\documentclass[prd,preprint,amsmath,amssymb,superscriptaddress,floatfix,nofootinbib,11pt]{revtex4-1}
\usepackage[utf8]{inputenc}
\usepackage{mathrsfs,bm}
\usepackage{natbib}
\usepackage{txfonts}
\usepackage{amssymb}
\usepackage{rotating}
\usepackage{indentfirst}
\usepackage{graphicx,booktabs}
\usepackage{multirow}
\usepackage{slashed}
\usepackage{overpic}
\usepackage{color}
\usepackage{amssymb}
\usepackage{hyperref}
\usepackage{bbding}
\usepackage{url}
\newcommand{\eL}{\epsilon_L}
\newcommand{\eR}{\epsilon_R}
\newcommand{\eSL}{\epsilon_{S_L}}
\newcommand{\eSR}{\epsilon_{S_R}}
\newcommand{\eT}{\epsilon_T}

\begin{document}
\title{Revisiting the new-physics interpretation of the $b\to c\tau\nu$ data}

\author{Rui-Xiang Shi}
\affiliation{School of Physics and
Nuclear Energy Engineering,  Beihang University, Beijing 100191, China}

\author{Li-Sheng Geng}
\email[E-mail: ]{lisheng.geng@buaa.edu.cn}
\affiliation{School of
Physics and Nuclear Energy Engineering \& International Research
Center for Nuclei and Particles in the Cosmos \& Beijing Key
Laboratory of Advanced Nuclear Materials and Physics,  Beihang
University, Beijing 100191, China}

\author{Benjam\'in Grinstein}
\affiliation{Department of Physics, University of California, San Diego, 9500 Gilman Drive, La Jolla, CA 92093-0319, USA}

\author{Sebastian J\"ager}
\affiliation{Department of Physics and Astronomy, University of Sussex, Brighton BN1 9QH, United Kingdom}

\author{Jorge Martin Camalich}
\affiliation{Instituto de Astrof\'isica de Canarias, C/ V\'ia L\'actea, s/n
E38205 - La Laguna, Tenerife, Spain
}
\affiliation{%
Universidad de La Laguna, Departamento de Astrof\'isica, La Laguna, Tenerife, Spain
}

\begin{abstract}

We revisit the status of the new-physics interpretations of the anomalies in semileptonic $B$ decays in light of the new data reported by Belle on the lepton-universality ratios $R_{D^{(*)}}$ using the semileptonic tag and on the longitudinal polarization of the $D^*$ in $B\to D^*\tau\nu$, $F_L^{D^*}$.
The preferred solutions involve new left-handed currents or tensor contributions. Interpretations with pure right-handed currents are disfavored by the LHC data, while pure scalar models are disfavored by the upper limits derived either from the LHC or from the $B_c$ lifetime. The observable $F_L^{D^*}$ also gives an important constraint leading to the exclusion of large regions of parameter space. Finally, we investigate the sensitivity of different observables to the various scenarios and conclude that a measurement of the tau polarization in the decay mode $B\to D\tau\nu$ would effectively discriminate among them.

\end{abstract}


\maketitle
\section{Introduction \label{sec:intro}}
For some time now, the ratios of semileptonic $B$-decay rates,
\begin{align}
R_{D^{(*)}}=\frac{{\rm BR}(B\to D^{(*)}\tau\nu)}{{\rm BR}(B\to D^{(*)}\ell\nu)}~~~~~~~~~\text{(with $\ell=e$ or $\mu$)},\label{eq:obs1}
\end{align}
have appeared to be enhanced with respect to the Standard Model (SM) predictions with a global significance above the evidence threshold~\cite{Lees:2012xj,Lees:2013uzd,Huschle:2015rga,Sato:2016svk,Aaij:2015yra,Hirose:2016wfn,Hirose:2017dxl,Aaij:2017uff,Aaij:2017deq,Aaij:2017tyk,Aoki:2016frl}.  In addition,  LHCb reports a value of the ratio
\begin{align}
R_{J/\psi}= \frac{{\rm BR}(B_c^+ \to J/\psi \tau^+ \nu_\tau)}{{\rm BR}(B_c^+ \to J/\psi \mu^+ \nu_\mu)},\label{eq:obs2}
\end{align}
about $2\sigma$ above the SM~\cite{Aaij:2017tyk}.

In the SM, semileptonic decays proceed via the tree-level exchange of a $W^\pm$ boson, preserving lepton universality. Hence, a putative NP contribution explaining the data must involve new interactions violating lepton universality. This may entail the tree-level exchange of new colorless vector ($W^\prime$)~\cite{Megias:2017ove,He:2017bft,Matsuzaki:2017bpp,Babu:2018vrl,Greljo:2018ogz,Asadi:2018wea} or scalar (Higgs)~\cite{Tanaka:1994ay,Celis:2012dk,Celis:2016azn,Iguro:2017ysu,Fraser:2018aqj,Martinez:2018ynq} particles, or leptoquarks~\cite{Sakaki:2013bfa,Alonso:2015sja,Barbieri:2015yvd,Freytsis:2015qca,Fajfer:2015ycq,Bauer:2015knc,Li:2016vvp,Barbieri:2016las,
Becirevic:2016yqi,Crivellin:2017zlb,Cai:2017wry,Assad:2017iib,DiLuzio:2017vat,Bordone:2017bld,
Altmannshofer:2017poe,Monteux:2018ufc,Marzocca:2018wcf,
Blanke:2018sro,Bordone:2018nbg,Becirevic:2018afm,Crivellin:2018yvo,Fornal:2018dqn,Angelescu:2018tyl,Baker:2019sli,Cornella:2019hct,Popov:2019tyc} with masses accessible to direct searches at the LHC.

Belle has also measured the longitudinal polarization of the $\tau$ ($P_{\tau}^{D^*}$)~\cite{Hirose:2016wfn} and of the $D^*$ ($F_L^{D^*}$)~\cite{Abdesselam:2019wbt} in the $B\to D^*\tau\nu$ decay,
\begin{eqnarray}
&&P_\tau^{D^*}=\frac{\Gamma(\lambda_\tau=\frac{1}{2})-\Gamma(\lambda_\tau=-\frac{1}{2})}
{\Gamma(\lambda_\tau=\frac{1}{2})+\Gamma(\lambda_\tau=-\frac{1}{2})},\nonumber\\
&&F_L^{D^*}=\frac{\Gamma(\lambda_{D^*}=0)}{\Gamma(\lambda_{D^*}=1)+\Gamma(\lambda_{D^*}=0)+\Gamma(\lambda_{D^*}=-1)},\label{eq:obs3}
\end{eqnarray}
where $\lambda_X$ refers to the helicity of the particle $X$. While $P_\tau^{D^*}$ is reconstructed from the hadronic decays of the $\tau$ and is still statistically limited, the reported measurement of $F_L^{D^*}$ is rather precise and disagrees with the SM prediction with a significance of $1.7\sigma$.

Recently, Belle announced a new combined measurement of both $R_D$ and $R_{D^*}$ using semileptonic decays for tagging the $B$ meson in the event~\cite{Abdesselam:2019dgh}. This presents a significant addition to the the data set because the previous combined measurements of $R_{D^{(*)}}$ had been performed at the $B$ factories using a hadronic tag. The new result is more consistent with the SM than the previous HFLAV average. Thus, these new data call for a reassessment of the significance of the tension of the signal with the SM and of the possible NP scenarios aiming at explaining it. The purpose of this work is to provide such an analysis using effective field theory (EFT)~\cite{Buttazzo:2017ixm,Alok:2017qsi,Azatov:2018knx,Bhattacharya:2018kig,Huang:2018nnq,Asadi:2018sym,Blanke:2018yud,Dutta:2017xmj,
Dutta:2013qaa,Hu:2018veh,Alok:2016qyh,Alok:2018uft} and to relate it to (partial) UV completions in terms of simplified mediators. We  assume that the lepton non-universal contribution affects only the couplings to the tau leptons. A comprehensive analysis of bounds on NP affecting $b \to c \ell \nu$ transitions can be found in ref.~\cite{Jung:2018lfu}. A summary of the recent  data (averages) is shown in Table~\ref{tab:1}, which is compared to the SM predictions which are obtained as specified in Sec.~\ref{sec:SM}.

\begin{table}[h!]
 \caption{
 Data (averages) and predictions in the SM for semileptonic b-decay observables defined in Eqs.\ref{eq:obs1}-\ref{eq:obs3}.
 The Heavy Flavor Averaging Group (HFLAV) 2018 averages~\cite{Amhis:2016xyh} of experimental data for $R_D$ and $R_{D^*}$ use data from BABAR~\cite{Lees:2012xj,Lees:2013uzd}, Belle~\cite{Huschle:2015rga,Sato:2016svk,Hirose:2016wfn} and LHCb~\cite{Aaij:2015yra,Aaij:2017uff,Aaij:2017deq}, while the HFLAV 2019 average includes the Belle measurement of both, $R_D$ and $R_{D^*}$, with the semileptonic tag~\cite{Abdesselam:2019dgh}. The LHCb measurement of $R_{J/\Psi}$ is reported in Ref.~\cite{Aaij:2017tyk} and the Belle measurements of $P_\tau^{D^*}$ and $F_L^{D^*}$ in Refs.~\cite{Aaij:2017tyk,Abdesselam:2019wbt}. The two experimental errors correspond to statistical and systematic uncertainties, respectively. SM predictions are obtained as specified in Sec.~\ref{sec:SM}.}\label{tab:1}
\begin{center}
    \begin{tabular}{ccccccc}
      \hline
      \hline
      Observables & \multicolumn{5}{c}{Data (averages)}& SM \\
      \hline
      \multirow{3}{*}{$R_D$}&\multicolumn{2}{c}{HFLAV 2018}&~&\multicolumn{2}{c}{HFLAV 2019}&\multirow{3}{*}{$0.312(19)$}\\
      \cline{2-3}\cline{5-6}
      &~~$0.407(39)(24)$&\multicolumn{1}{c}{}&~&~~$0.340(27)(13)$&&\\
     \cline{1-1}\cline{7-7}
      $R_{D^*}$&~~$0.306(13)(7)$&\multicolumn{1}{c}{\raisebox{0.5cm}{~~~$\text{corr}=-0.20$~~~}}&~&~~$0.295(11)(8)$&\raisebox{0.5cm}{~~~$\text{corr}=-0.38$~~~}&$0.253(4)$\\
      \hline
      $R_{J/\psi}$&\multicolumn{5}{c}{$0.71(17)(18)$}&$0.248(3)$ \\
      \hline
      $P_\tau^{D^*}$&\multicolumn{5}{c}{$-0.38(51)(19)$}&$-0.505(23)$ \\
      \hline
    $F_L^{D^*}$&\multicolumn{5}{c}{$0.60(8)(4)$}&$0.455(9)$ \\
    \hline
    \hline
    \end{tabular}
  \end{center}
\end{table}

\section{Theoretical framework}
\label{sec:TH}
\subsection{Low-energy effective Lagrangian}
\label{sec:EFT}

The most general effective Lagrangian describing the contributions of heavy NP to semitauonic $b\to c\tau\bar\nu$ processes can be written as
\begin{eqnarray}
\label{eq:EFTLag}
{\cal L}_{\rm eff}^{\rm LE}\supset&&-\frac{4G_F V_{cb}}{\sqrt{2}}[(1+\epsilon_L^\tau)(\bar{\tau}\gamma_\mu P_L \nu_\tau)(\bar{c}\gamma^\mu P_Lb)+\epsilon_R^\tau(\bar{\tau}\gamma_\mu P_L\nu_\tau)(\bar{c}\gamma^\mu P_Rb)\nonumber\\
&&+\epsilon_{S_L}^\tau(\bar{\tau}P_L\nu_\tau)(\bar{c}P_Lb)+\epsilon_{S_R}^\tau(\bar{\tau}P_L\nu_\tau)(\bar{c}P_Rb)+\epsilon_T^\tau(\bar{\tau}\sigma_{\mu\nu} P_L\nu_\tau)(\bar{c}\sigma^{\mu\nu}P_Lb)]+\mbox{H.c.},
\end{eqnarray}
where $G_F$ is the Fermi constant and $V_{cb}$ is the Cabibbo-Kobayashi-Maskawa (CKM) matrix element. The five Wilson coefficients (WCs) $\epsilon_L^\tau$, $\epsilon_R^\tau$, $\epsilon_T^\tau$, $\epsilon_{S_L}^\tau$ and $\epsilon_{S_R}^\tau$ encapsulate the NP contributions, featuring the scaling $\epsilon_\Gamma^\tau\sim\mathcal O(v^2/\Lambda_{\rm NP}^2)$, where $v\approx246$~GeV is the electroweak symmetry  breaking (EWSB) scale. In the context of the  EFT of the SM (SMEFT)~\cite{Buchmuller:1985jz,Grzadkowski:2010es}, $\epsilon_R^\tau=\epsilon_R^\ell+{\cal O}(v^4/\Lambda_{\rm NP}^4)$ and the right-handed operator cannot contribute to lepton universality violation at leading order in the $(v^2/\Lambda_{\rm NP}^2)$ expansion~\cite{Bernard:2006gy,Cirigliano:2009wk,Alonso:2015sja}. For this reason, we do not consider the effect of $\epsilon_R^\tau$ in our fits.  Nonetheless, it is important to note that this assumption could be relaxed if there was not a mass gap between the NP and the EWSB scales, or under a nonlinear realization of the electroweak symmetry breaking~\cite{Cata:2015lta}.

The chirally-flipping scalar and tensor operators are renormalized by QCD and electroweak corrections~\cite{Gonzalez-Alonso:2017iyc,Aebischer:2017gaw,Jenkins:2017dyc,Feruglio:2018fxo}. The latter induce a large mixing of the tensor operator into $\epsilon^{\tau}_{S_L}$ which can have relevant implications for tensor scenarios~\cite{Gonzalez-Alonso:2017iyc}. As an illustration, defining $\vec \epsilon^{~T}(\mu)=(\epsilon_{S_R}^\tau,\epsilon_{S_L}^\tau,\epsilon_{T}^\tau)(\mu)$, (where we have omitted flavor indices), we find that $\vec \epsilon~(m_b)=M~\vec \epsilon~(1~\text{TeV})$, with~\cite{Gonzalez-Alonso:2017iyc}
\begin{align}
\label{eq:RGEepsilon}
M=
\left(
\begin{array}{ccc}
1.737 &0 & 0 \\
0& 1.752 & -0.287\\
0& -0.0033 & 0.842 \\
\end{array}
\right),
\end{align}
and where, in a slight abuse of notation, we keep the notation for the WCs of the low-energy EFT above the EWSB scale.
Operators with vector currents do not get renormalized by QCD, whereas electromagnetic and electroweak corrections produce a correction of a few percent to the tree-level contributions~\cite{Sirlin:1981ie,Gonzalez-Alonso:2017iyc}. On the other hand, all the operators in the SMEFT matching at low-energies to the Lagrangian in eq.~(\ref{eq:EFTLag}) can give, under certain assumptions on the flavor structure of the underlying NP, large contributions to other processes such as decays of electroweak bosons, the $\tau$ lepton and the Higgs, or the anomalous magnetic moment of the muon~\cite{Feruglio:2016gvd,Feruglio:2017rjo,Feruglio:2018fxo}.

An interesting scenario where the new physics cannot be described by the local effective Lagrangian
eq.~(\ref{eq:EFTLag}) consists of the addition of new light right-handed neutrinos~\cite{Becirevic:2016yqi,He:2017bft,Greljo:2018ogz,Asadi:2018wea,Babu:2018vrl,Robinson:2018gza,Azatov:2018kzb}. This duplicates the operator basis given in eq.~(\ref{eq:EFTLag}) by the replacements $P_L\to P_R$ in the leptonic currents (and in the hadronic current for the tensor operator)~\cite{Goldberger:1999yh,Cirigliano:2009wk,Robinson:2018gza} and whose WCs we label with $\epsilon_\Gamma\to\tilde\epsilon_\Gamma$. None of these operators interfere with the SM and their contributions to the decay rates are, thus, quadratic and positive. This also means that the size of the NP contributions needed to explain $R_{D^{(*)}}$ in this case are larger than with the operators in eq.~(\ref{eq:EFTLag}) and they typically enter in conflict with bounds from other processes like the decay $B_c\to\tau\nu$~\cite{Alonso:2016oyd,Akeroyd:2017mhr} or from direct searches at the LHC~\cite{Greljo:2018tzh}. As an illustration of the features and challenges faced by these models we consider the operator with right-handed currents,
\begin{eqnarray}
\label{eq:EFTLag2}
{\cal L}_{\rm eff}^{\rm LE}\supset-\frac{4G_F V_{cb}}{\sqrt{2}}(\tilde\epsilon_R^\tau\bar{\tau}\gamma_\mu N_R)(\bar{c}\gamma^\mu P_Rb)+\mbox{H.c.},
\end{eqnarray}
(with $N_R$ denoting the right-handed neutrino), which incarnates a popular NP interpretation of the anomaly~\cite{He:2017bft,Greljo:2018ogz,Asadi:2018wea,Babu:2018vrl,Robinson:2018gza,Azatov:2018kzb}. Finally, imaginary parts also contribute quadratically to the rates so we assume the WCs to be real, although we will briefly study also the impact of imaginary parts below.

\subsection{Simplified models}
\label{sec:mediators}

\begin{table}[h!]
\caption{Quantum numbers of mediators that can explain at tree-level the $R_{D^{(*)}}$ anomalies and their contributions to the effective operators in eqs.~(\ref{eq:EFTLag}),~(\ref{eq:EFTLag2}).\label{tab:mediators}}
\begin{center}
    \begin{tabular}{cccccccccc}
     \hline
     \hline
     Mediator&Spin&$S\!U(3)$&$S\!U(2)$&$U(1)$&$\eL^\tau$&$\tilde\epsilon_R^\tau$&$\eSR^\tau$&$\eSL^\tau$& $\eT^\tau$\\
     \hline
     $H$&0& {\bf 1} & {\bf 2} & $+1/2$ &{\color{red}\XSolidBrush}&{\color{red}\XSolidBrush} &{\color{green}\CheckmarkBold} &{\color{green}\CheckmarkBold} & {\color{red}\XSolidBrush}\\
     $W^{\prime}_L$& 1& {\bf 1} & {\bf 3} & 0 &{\color{green}\CheckmarkBold} & {\color{red}\XSolidBrush}& {\color{red}\XSolidBrush}&{\color{red}\XSolidBrush}& {\color{red}\XSolidBrush}\\
      $W^{\prime}_R$&1& {\bf 1} & {\bf 1} & +1 & {\color{red}\XSolidBrush} &{\color{green}\CheckmarkBold} & {\color{red}\XSolidBrush}&{\color{red}\XSolidBrush}& {\color{red}\XSolidBrush}\\
     \hline
     $S_1$ &0& ${\boldsymbol{\overline 3}}$ & {\bf 1} & +1/3 &{\color{green}\CheckmarkBold} &{\color{green}\CheckmarkBold}& {\color{red}\XSolidBrush}&{\color{green}\CheckmarkBold} &{\color{green}\CheckmarkBold}\\
     $S_3$ & 0& ${\boldsymbol{\overline 3}}$ & {\bf 3} & +1/3 &{\color{green}\CheckmarkBold} &{\color{green}\CheckmarkBold}& {\color{red}\XSolidBrush}&{\color{red}\XSolidBrush} &{\color{red}\XSolidBrush}\\
     $R_2$ & 0& {\bf 3} & {\bf 2} & +7/6& {\color{green}\CheckmarkBold}&{\color{green}\CheckmarkBold}& {\color{red}\XSolidBrush}&{\color{green}\CheckmarkBold}  &{\color{green}\CheckmarkBold}\\

     $U_1$ & 1& {\bf 3} & {\bf 1} & +2/3 &{\color{green}\CheckmarkBold} &{\color{green}\CheckmarkBold}& {\color{green}\CheckmarkBold}& {\color{red}\XSolidBrush}& {\color{red}\XSolidBrush} \\
     $U_3$ & 1 &{\bf  3} & {\bf 3} & +2/3 &{\color{green}\CheckmarkBold} &{\color{green}\CheckmarkBold}& {\color{red}\XSolidBrush}&{\color{red}\XSolidBrush} &{\color{red}\XSolidBrush}\\
     $V_2$ & 1 &${\boldsymbol{\overline 3}}$& {\bf 2} & +5/6 &{\color{red}\XSolidBrush} &{\color{red}\XSolidBrush}& {\color{green}\CheckmarkBold}&{\color{red}\XSolidBrush} &{\color{red}\XSolidBrush}\\
     \hline
     \hline
    \end{tabular}
\end{center}
\end{table}

The effective operators in eqs.~(\ref{eq:EFTLag}),~(\ref{eq:EFTLag2}) can be mediated at tree level by a number of new particles, that we list in Tab.~\ref{tab:mediators}. Possibilities with new charged colorless weak bosons can be realized with the $W^\prime$ in either a triplet ($W^\prime_L$) or a singlet ($W^\prime_R$) representation of weak isospin. In the former case, the neutral component of the triplet, a $Z^\prime$ with a mass close to the one of the $W^\prime$, produces large effects in either neutral-meson mixing or di-tau production at the LHC, so that this scenario is unavoidably in conflict with data~\cite{Faroughy:2016osc}. Making the $W^\prime$ a singlet of weak isospin, $W^\prime_R$=({\bf 1}, {\bf 1},+1) under $S\!U(3)\times S\!U(2)\times U(1)$, requires introducing right-handed neutrinos to contribute to $b\to c\tau\bar\nu$ ~\cite{He:2017bft,Greljo:2018ogz,Asadi:2018wea,Babu:2018vrl}; parametrizing the Lagrangian for this model,
\begin{align}
\label{eq:LagWp}
\mathcal L_{W^\prime} \supset \left(g_{cb}\bar c\gamma_\mu P_R b+g_{\tau N}\bar N_R\gamma_\mu P_R\tau\right)W^{\prime\mu}_R+{\rm h.c.},
\end{align}
one finds the contribution to the EFT,
\begin{align}
\label{eq:LagEFTWp}
V_{cb}\tilde \epsilon_R^\tau=\frac{g_{cb}g_{\tau N}^*}{2}\frac{v^2}{m_{W^\prime}^2}.
\end{align}
Models based on extending the scalar sector of the SM, such as the two-Higgs doublet model (labeled by $H$ in Tab.~\ref{tab:mediators}), generate the scalar operators through charged-Higgs exchange. However, these are disfavored by experimental bounds that stem from the $B_c$ lifetime~\cite{Alonso:2016oyd} and from the branching fraction of $B_c\to \tau\nu$ derived using LEP data~\cite{Akeroyd:2017mhr}. Strong limits from direct searches at the LHC of the corresponding charged scalars have also been obtained in the literature~\cite{Iguro:2018fni}.

On the other hand, leptoquark exchanges can produce all the operators in eq.~(\ref{eq:EFTLag}).~\footnote{We follow the notation to label the leptoquark fields introduced in refs.~\cite{Buchmuller:1986zs,Dorsner:2016wpm}.} The SM interactions of the scalar leptoquark $S_1$=(${\boldsymbol{\overline 3}}$,~{\bf1},+1/3) can be described by the Lagrangian,
\begin{align}
\label{eq:LagS1}
\mathcal L_{S_1}\supset y_{1,i\alpha}^{LL}\bar Q_{L,i}^c~\epsilon~L_{L,\alpha}S_1+y_{1,i\alpha}^{RR}\bar u_{R,i}^c~e_{R,\alpha}S_1+y_{1,i\alpha}^{\overline{RR}}\bar d_{R,i}^c~N_{R,\alpha}S_1,
\end{align}
where $\epsilon_{ab}$ is the antisymmetric tensor of rank two and where we are labeling the flavor of the fields in the interaction basis. This model produces left-handed, scalar-tensor and right-handed contributions~\cite{Sakaki:2013bfa,Bauer:2015knc,Cai:2017wry,Crivellin:2017zlb},
\begin{align}
\label{eq:LagEFTS1}
V_{cb}\eL^{\tau}=\frac{\tilde y_{1,33}^{LL,d}(\tilde y_{1,23}^{LL,u})^*}{4}\frac{v^2}{m_{S_1}^2},~~~V_{cb}\eSL^{\tau}=-4V_{cb}\eT^{\tau}=\frac{\tilde y_{1,33}^{LL,d}(\tilde y_{1,23}^{RR})^*}{4}\frac{v^2}{m_{S_1}^2},~~~V_{cb}\tilde\epsilon_R^{\tau}=-\frac{\tilde y_{1,33}^{\overline{RR}}(\tilde y_{1,23}^{RR})^*}{4}\frac{v^2}{m_{S_1}^2},
\end{align}
where the coefficients are defined at a scale equal to the leptoquark mass, $\mu=m_{S_1}$. The tilde in the coefficients of eq.~(\ref{eq:LagEFTS1}) and in the rest of this subsection indicates that the quark unitary rotations have been absorbed in the definition of the couplings. For instance, if such transformations are $d_L\to L_d~d_L$, $u_L\to L_u~u_L$, $d_R\to R_d~d_R$, $u_R\to R_u~u_R$, we have defined $\tilde y_{1,i\alpha}^{LL,u(d)}=[y_{1}^{LL}~L_{u(d)}]_{i\alpha}$, $\tilde y_{1,i\alpha}^{RR}=[y_{1}^{RR}~R_u]_{i\alpha }$ and $\tilde y_{1,i\alpha}^{\overline{RR}}=[y_{1}^{\overline{RR}}~R_d]_{i\alpha}$ where summation of quark flavor indices is implicit. We have also defined these couplings in the charged-lepton mass basis, ignoring neutrino masses.

 The leptoquark with quantum numbers $R_2$=({\bf 3},~{\bf 2},+7/6) and Lagrangian,
 \begin{align}
 \label{eq:R2}
 \mathcal L_{R_2}\supset -y_{2,i\alpha}^{RL}\bar u_{R,i}~\epsilon~L_{L,\alpha}R_2+y_{2,i\alpha}^{LR}\bar Q_{L,i} e_{R,\alpha}R_2,
 \end{align}
 leads to
 \begin{align}
 \label{eq:LagEFTR2}
 V_{cb}\eSL^{\tau}=+4V_{cb}\eT^{\tau}=\frac{\tilde y_{2,23}^{RL}(\tilde y_{2,33}^{LR,d})^*}{4}\frac{v^2}{m_{R_2}^2}.
 \end{align}
 Thus, one can achieve a tensor scenario by adjusting the masses and couplings of the $S_1$ and $R_2$ leptoquarks. It is important to stress that such a solution at low energies requires some tuning due to the large electroweak mixing into scalar operators in eq.~(\ref{eq:RGEepsilon}).

Among the the vector leptoquarks we consider the $U_1$=({\bf 3}, {\bf 1},+2/3), which has been extensively studied in the interpretation of the $B$ anomalies~\cite{Alonso:2015sja,Barbieri:2015yvd,Assad:2017iib,DiLuzio:2017vat,Bordone:2017bld,Monteux:2018ufc,Marzocca:2018wcf,Blanke:2018sro,Bordone:2018nbg,Crivellin:2018yvo,Angelescu:2018tyl,Baker:2019sli,Cornella:2019hct},
 \begin{align}
  \mathcal L_{U_1}\supset \chi_{1,i\alpha}^{LL} \bar Q_{L,i}\gamma_\mu L_{L,\alpha}U_1^\mu+\chi_{1,\alpha}^{RR} \bar d_{R,i}\gamma_\mu e_{R,\alpha}U_1^\mu+
  \chi_{1,i\alpha}^{\overline{RR}} \bar u_{R,i}\gamma_\mu N_{R,\alpha}U_1^\mu,
 \end{align}
 leading to left-handed and right-handed contributions, and a scalar contribution,
 \begin{align}
 \label{eq:LagEFTU1}
 V_{cb}\eL^{\tau}=\frac{\tilde \chi_{1,23}^{LL,u}(\tilde \chi_{1,33}^{LL,d})^*}{2}\frac{v^2}{m_{U_1}^2},~~~V_{cb}\tilde\epsilon_R^{\tau}=\frac{\tilde \chi_{1,23}^{\overline{RR}}(\tilde \chi_{1,33}^{RR})^*}{2}\frac{v^2}{m_{U_1}^2},~~~V_{cb}\epsilon_{S_R}^\tau=-\tilde\chi_{1,23}^{LL,u}(\chi_{1,33}^{\overline{RR}})^*\frac{v^2}{m_{U_1}^2}.
\end{align}
In particular, a combination of left-handed and right-handed couplings gives rise to a scalar operator which is instrumental to achieve a better agreement with data in some UV completions of the $U_1$ leptoquark~\cite{Bordone:2017bld,Bordone:2018nbg,Baker:2019sli,Cornella:2019hct}.

The mediators $S_3$=(${\boldsymbol{\overline 3}}$,~{\bf3},+1/3) and $U_3$=(${\boldsymbol{\overline 3}}$,~{\bf3},+2/3) in Tab.~\ref{tab:mediators} provide completions of the left-handed current operator equivalent to the $S_1$ and $U_1$ ones for scalar and vector leptoquark scenarios, respectively. Finally, we have not included in the table the leptoquarks $\tilde R_2=({\boldsymbol 3}$,~{\bf2}, $+1/6$) and $\tilde V_2=({\boldsymbol{\overline 3}}$,~{\bf2}, $-1/6$) because they only contribute to scalar and tensor operators with right-handed neutrinos which are not considered in this work, as argued in Sec.~\ref{sec:EFT}.

\subsection{Form factors}
\label{sec:SM}

The hadronic matrix elements in the $b\rightarrow c$ decay amplitudes are parameterized in terms of the following form factors,
\begin{eqnarray}
&&\langle D(k)|\bar{c}\gamma^\mu b|\bar{B}(p)\rangle=(p+k)^\mu f_+(q^2)+(p-k)^\mu\frac{m_B^2-m_D^2}{q^2}(f_0(q^2)-f_+(q^2)),~~~\langle D(k)|\bar{c} b|\bar{B}(p)\rangle=\frac{m_B^2-m_D^2}{m_b-m_c}f_0(q^2),\nonumber\\
&&\langle D(k)|\bar{c}\sigma^{\mu\nu}b|\bar{B}(p)\rangle=\frac{2if_T(q^2)}{m_B+m_D}(k^\mu p^\nu-p^\mu k^\nu),~~~~\langle D(k)|\bar{c}\sigma^{\mu\nu}\gamma_5b|\bar{B}(p)\rangle=\frac{2f_T(q^2)}{m_B+m_D}\epsilon^{\mu\nu\alpha\beta}k_\alpha p_\beta,\nonumber
\end{eqnarray}
\begin{eqnarray}
\langle V(k,\epsilon)|\bar{c}\gamma^\mu b|P(p)\rangle=\frac{2iV(q^2)}{m_P+m_V}\epsilon^{\mu\nu\alpha\beta}\epsilon_{\nu}^*k_\alpha p_\beta,~~~\langle V(k,\epsilon)|\bar{c}\gamma_5b|P(p)\rangle=-\frac{2m_V}{m_b+m_c}A_0(q^2)\epsilon^*\cdot q,\nonumber
\end{eqnarray}
\begin{eqnarray}
\langle V(k,\epsilon)|\bar{c}\gamma^\mu\gamma_5 b|P(p)\rangle&=&2m_VA_0(q^2)\frac{\epsilon^*\cdot q}{q^2}q^\mu+(m_P+m_V)A_1(q^2)\left(\epsilon^{*\mu}-\frac{\epsilon^*\cdot q}{q^2}q^\mu\right)\nonumber\\
&&-A_2(q^2)\frac{\epsilon^*\cdot q}{m_P+m_V}\left((p+k)^\mu-\frac{m_P^2-m_V^2}{q^2}q^\mu\right)\nonumber
\end{eqnarray}
\begin{eqnarray}
\langle V(k,\epsilon)|\bar{c}\sigma^{\mu\nu}b|P(p)\rangle&=&\frac{\epsilon^*\cdot q}{(m_P+m_V)^2}T_0(q^2)\epsilon^{\mu\nu\alpha\beta}p_\alpha k_\beta+T_1(q^2)\epsilon^{\mu\nu\alpha\beta}p_\alpha\epsilon_{\beta}^*+T_2(q^2)\epsilon^{\mu\nu\alpha\beta}k_\alpha\epsilon_{\beta}^*,\nonumber\\
\langle V(k,\epsilon)|\bar{c}\sigma^{\mu\nu}\gamma_5b|P(p)\rangle&=&\frac{i\epsilon^*\cdot q}{(m_P+m_V)^2}T_0(q^2)(p^\mu k^\nu-k^\mu p^\nu)\nonumber\\
&&+iT_1(q^2)(p^\mu\epsilon^{*\nu}-\epsilon^{*\mu}p^\nu)+iT_2(q^2)(k^\mu\epsilon^{*\nu}-\epsilon^{*\mu}k^\nu),
\end{eqnarray}
where $q=p-k$, $\epsilon_{0123}=1$, $V$ and $P$ stand for vector mesons  ($D^*$ and $J/\psi$)  and pseudoscalar mesons
($B$ and $B_c$), respectively. We take the quark masses in the $\overline{\rm MS}$ scheme, i.e, $m_b\equiv\overline{m}_b(\overline{m}_b)=4.18$~GeV and $\overline{m}_c(\overline{m}_c)=1.27$~GeV~\cite{Tanabashi:2018oca},. Note that the $c$-quark mass is derived by the solution of the renormalization group equation for $\overline{m}_c(\mu)$ at two-loop order and $\alpha_s(\mu)$ with three-loop accuracy~\cite{Buchalla:1995vs}. We follow the PDG~\cite{Tanabashi:2018oca} for the masses of the mesons relevant in this work.

For the $B\to D^{(*)}$ mode, some of the form factors are taken from  Lattice QCD calculations~\cite{Na:2015kha,Bailey:2014tva}. The rest are parameterized using heavy-quark effective theory (HQET)~\cite{Shifman:1987rj,Isgur:1989vq,Isgur:1989ed,Falk:1990yz,Boyd:1994tt,Boyd:1995cf,Caprini:1997mu,Fajfer:2012vx} whose nuisance parameters are determined by the HFLAV global fits to the $\bar{B}\to D^{(*)}\ell^-\bar{\nu}$ data~\cite{Amhis:2014hma}. Our determination of $R_D$ and $R_{D^*}$ differs from that of HFLAV in the choice of form factors; ours, based on Ref.~\cite{Alonso:2016gym}, do not include some recent refinements~\cite{Bernlochner:2017jka,Bigi:2017jbd,Jaiswal:2017rve}.

For the $B_c\to J/\psi$ form factors, they have been studied in a variety of approaches~\cite{Wen-Fei:2013uea,Kiselev:2002vz,Fu:2018vap,Zhu:2017lqu,Shen:2014msa,Wang:2008xt,Hernandez:2006gt,Ebert:2003cn,Lytle:2016ixw,
Colquhoun:2016osw,Tran:2018kuv} (for earlier analysis focused on this decay mode see refs.~\cite{Watanabe:2017mip,Dutta:2017wpq,Tran:2018kuv,Cohen:2018dgz}). Here we take $V(q^2)$, $A_0(q^2)$, $A_1(q^2)$ and $A_2(q^2)$ calculated in the covariant light-front quark model~\cite{Wang:2008xt} because these results are well consistent with the lattice results at all available $q^2$ points in Ref.~\cite{Lytle:2016ixw,
Colquhoun:2016osw}. The three tensor form factors can be related through the corresponding HQET form factor $h_{A_1}(\omega)$ at leading order in the heavy-quark expansion,
\begin{eqnarray}
&&A_1(\omega)=(\omega+1)\frac{\sqrt{m_{B_c}m_{J/\psi}}}{m_{B_c}+m_{J/\psi}}h_{A_1}(\omega),~~T_0(\omega)={\cal O}(\Lambda/m_Q),\nonumber\\
&&T_1(\omega)=\sqrt{m_{J/\psi}/m_{B_c}}h_{A_1}(\omega)+{\cal O}(\Lambda/m_Q),~T_2(\omega)=\sqrt{m_{B_c}/m_{J/\psi}}h_{A_1}(\omega)+{\cal O}(\Lambda/m_Q),
\end{eqnarray}
with $\omega=\upsilon_{J/\psi}\cdot\upsilon_{B_c}=(m_{B_c}^2+m_{J/\psi}^2-q^2)/(2m_{B_c}m_{J/\psi})$ and
where we have neglected the $\Lambda/m_Q$ power corrections.

\subsection{Statistical Method}
\label{sec:statistics}

We follow a frequentist statistical approach to compare the measured values of
   $n_{\rm exp}$ observables, $\vec{O}^{\rm exp}$, to their  theoretical predictions $\vec{O}^{\rm th}$ as functions
of the Wilson coefficients $\vec \epsilon$, and of nuisance theoretical parameters $\vec y$.
The nuisance parameters parameterize the lack of knowledge (theoretical uncertainties)
of the form factors. For the $B\to D^{(*)}$ decays, we employ the
parametrization and numerical inputs (including correlations) described in ref.~\cite{Alonso:2016gym}. For the $B_c\to J/\Psi$ decays we parameterize the theoretical errors reported for the form factors in ref.~\cite{Wang:2008xt} as uncorrelated nuisance parameters. We then define a test statistic
\begin{eqnarray}
\tilde \chi^2(\vec\epsilon,~\vec y)&=&\chi_{\rm exp}^2(\vec\epsilon,~\vec y)+\chi_{\rm th}^2(\vec y),
\end{eqnarray}
where
\begin{eqnarray}
&&\chi_{\rm exp}^2(\vec \epsilon,~\vec y)=[\vec O^{~\rm th}(\vec \epsilon,~\vec y)-\vec O^{~\rm exp}]^T\cdot (V^{\rm exp})^{-1}\cdot [\vec O^{~\rm th}(\vec \epsilon,~\vec y)-\vec O^{~\rm exp}],\nonumber\\
&&\chi_{\rm th}^2(\vec y)=(\vec y-\vec y_0)^T\cdot(V^{\rm th})^{-1}\cdot(\vec y-\vec y_0),
\end{eqnarray}
$\vec y_0$ are a set of central values for the nuisance parameters,
and $V^{\rm exp}$ and $V^{\rm th}$ denote the experimental and theoretical covariance matrices, respectively. By
adding the theory term $\chi^2_{\rm th}$ we have in effect (from a statistical point of view) added
$n_{\rm th}$ (correlated) ``measurements''  of the $n_{\rm th}$ theory parameters to the $n_{\rm exp}$ measurements
of the observables.

We will consider scenarios (statistical models) with different subsets of the Wilson coefficients allowed to vary and the remaining ones set to zero,
and with various subsets of the experimental observables included. In each case, we
obtain best-fit values for the model parameters, including the nuisance parameters, by minimizing $\chi^2$. To do so,
in a first step we construct a profile-$\chi^2$ function
\begin{eqnarray}
\chi^2(\vec \epsilon)=\underset{\vec{y}}{\rm min}~\tilde \chi^2(\vec\epsilon,~\vec y),
\end{eqnarray}
which depends solely on the subset of Wilson coefficients allowed to take nonzero values in a particular scenario, which we
again refer to as $\vec \epsilon$. (Note that in the case of a single
measurement of an observable whose theoretical expression depends linearly on a single theory nuisance parameter $y$, such
that $y-y_0$ is proportional to the theoretical uncertainty, the profiling reproduces the widely employed
prescription of combining theoretical and experimental errors in quadrature.)
In a second step, we minimize $\chi^2(\vec \epsilon)$ over $\vec \epsilon$;
the value(s) of $\vec \epsilon$ at the minimum $\chi^2_{\rm min}$
provide(s) the best fit (maximum likelihood fit).

Next, we compute a $p$-value to quantify the goodness of fit, i.e.\ how well a given scenario can describe the data.
We will assume that $\chi^2(\vec \epsilon)$ follows a $\chi^2$-distribution with $n_{\rm dof} = n_{\rm exp} - n_\epsilon$ degrees of freedom, where $n_\epsilon$ is the number of parameters allowed to vary in a given fit.
Note that the theory parameters do not contribute to $n_{\rm dof}$ because $\chi^2_{\rm th}$
contains as many ``measurements'' as theory parameters. In each scenario, the $p$-value is obtained from
$\chi^2_{\rm min}$ as one minus the cumulative $\chi^2$ distribution for $n_{\rm dof}$ degrees of freedom.
To illustrate this, let us consider only the $\chi^2_{\rm exp}$ including $R_D$ and $R_{D*}$
and ask how well the SM describes these data. For simplicity, let us neglect
theory errors altogether (they will be included in the following section, with little impact on the result), taking
the SM prediction to be the central values employed by HFLAV2019, $R_D^{\rm SM, HFLAV} = 0.299$
and $R_{D^*}^{\rm SM, HFLAV} = 0.258$. In this case, there are no parameters to minimize over
and $\chi^2$ is simply a number. This is easily obtained from
the HFLAV2019 averages and correlation shown in Table~\ref{tab:1}, substituting the SM values for the observables, which gives $\Delta \chi^2_\mathrm{SM} = \Delta \chi^2(\vec 0)$ as defined below,
and adding a constant $\chi^2_{\rm min} = 8.7$ as stated
by HFLAV~\footnote{By adding $\chi^2_{\rm min}$ we are taking into account the goodness of the HFLAV fit to the different measurements of $R_{D^{(*)}}$ which is needed to obtain an accurate estimate of the $p$-values.}.
 Nine measurements entered the combination and we are determining
zero parameters, resulting in $n_{\rm dof} = 9$. With $\chi^2_{\rm SM} = \chi^2(\vec \epsilon= \vec 0) = 22.8$,
this gives a  $p$-value of $6.56 \times 10^{-3}$ corresponding to $2.72 \sigma$,
slightly reduced from $3.00 \sigma$ obtained in an analogous manner from the HFLAV2018 combination.

Finally, for each one-parameter BSM scenario, we construct
$\Delta \chi^2(\vec \epsilon) = \chi^2(\vec \epsilon) - \chi^2_{\rm min}$ and
obtain $n\sigma$  confidence intervals from the requirement $\Delta \chi^2 \leq n^2$.
Similarly, for each 2-parameter scenario we construct the corresponding $\Delta \chi^2$ and obtain
two-dimensional $1\sigma$ and $2\sigma$ regions from the conditions $\Delta \chi^2 \leq 2.3$
and $\Delta \chi^2 \leq 6.18$, respectively. We also determine, for each model,
$$
\Delta \chi^2_\mathrm{SM} = \chi^2(\vec 0) - \chi^2_{\rm min},
$$
to quantify at what level the SM point is excluded in that model. The $\sqrt{\Delta \chi^2_\mathrm{SM}}$ is converted to an equivalent number
of standard deviations, referred to as the pull $\mathrm{Pull}_\mathrm{SM}$,
by employing the cumulative $\chi^2$-distribution with $n_{\rm dof}$ set to 1 or 2, the number of jointly determined
parameters, as appropriate.

Let us close this section by contrasting to the usual approach for comparing the
$R_D$ and $R_{D^*}$ measurements to the SM, as employed by HFLAV. In this approach, the true values
of $R_D$ and $R_{D^*}$ are treated as free parameters, which effectively amounts to a two-parameter
BSM model. In this model, HFLAV obtain an SM pull of $3.08\sigma$. We stress that this is a statement about how
much better than the SM a BSM model can potentially describe the data. It is conceptually analogous to the pulls
in our two-parameter Wilson coefficient fits.
(In fact, we will find in the next section a slightly higher pull for two of our 1-parameter models. This comes about
because a given $\Delta \chi^2$ value implies a lower $p$-value (higher number of standard deviations) when determining
a single parameter as  opposed to joint determination of two parameters.)
Conversely, our SM $p$-values are a statement how well the SM describes the data, \textit{without reference to
any comparator BSM model}. As we have seen, the data is marginally consistent with the SM at $3\sigma$, little
changed from 2018. As we will see in the subsequent sections, the impact of the new Belle data on the best-fit
values in BSM scenarios is much stronger.

\section{Results}
\label{sec:Results}

In this section, we investigate the values of the WCs determined by fitting to the experimental data of  $R_D$, $R_{D^*}$, $R_{J/\psi}$, $P_\tau^{D^*}$ and $F_L^{D^*}$ given in Table~\ref{tab:1}. We also discuss the constraints on scalar operators derived from the limits ${\rm Br}(B_c\to \tau\nu)\leq30\%(10\%)$ which are obtained using the $B_c$ lifetime~\cite{Alonso:2016oyd} (LEP searches of the decays $B_{(c)}\to\tau\nu$~\cite{Akeroyd:2017mhr}). Note that these limits have been critically discussed in Refs.~\cite{Blanke:2018yud,Blanke:2019qrx,Bardhan:2019ljo}. Finally, an upper bound on the values of the WCs can be derived from the tails of the monotau signature ($p p\to \tau_h X$+MET) at the LHC~\cite{Greljo:2018tzh,Aaboud:2018vgh,Sirunyan:2018lbg} (see below). We will perform fits to two types of dataset: $R_{D^{(*)}}$ only, as well as to the full dataset in Table~\ref{tab:1} including in addition $R_{J/\psi}$ and
the polarization observables.

\subsection{Fits to $R_{D^{(*)}}$ only}
\label{sec:fitsRDs}

\begin{figure}[h!]
\centering
\includegraphics[width=8cm]{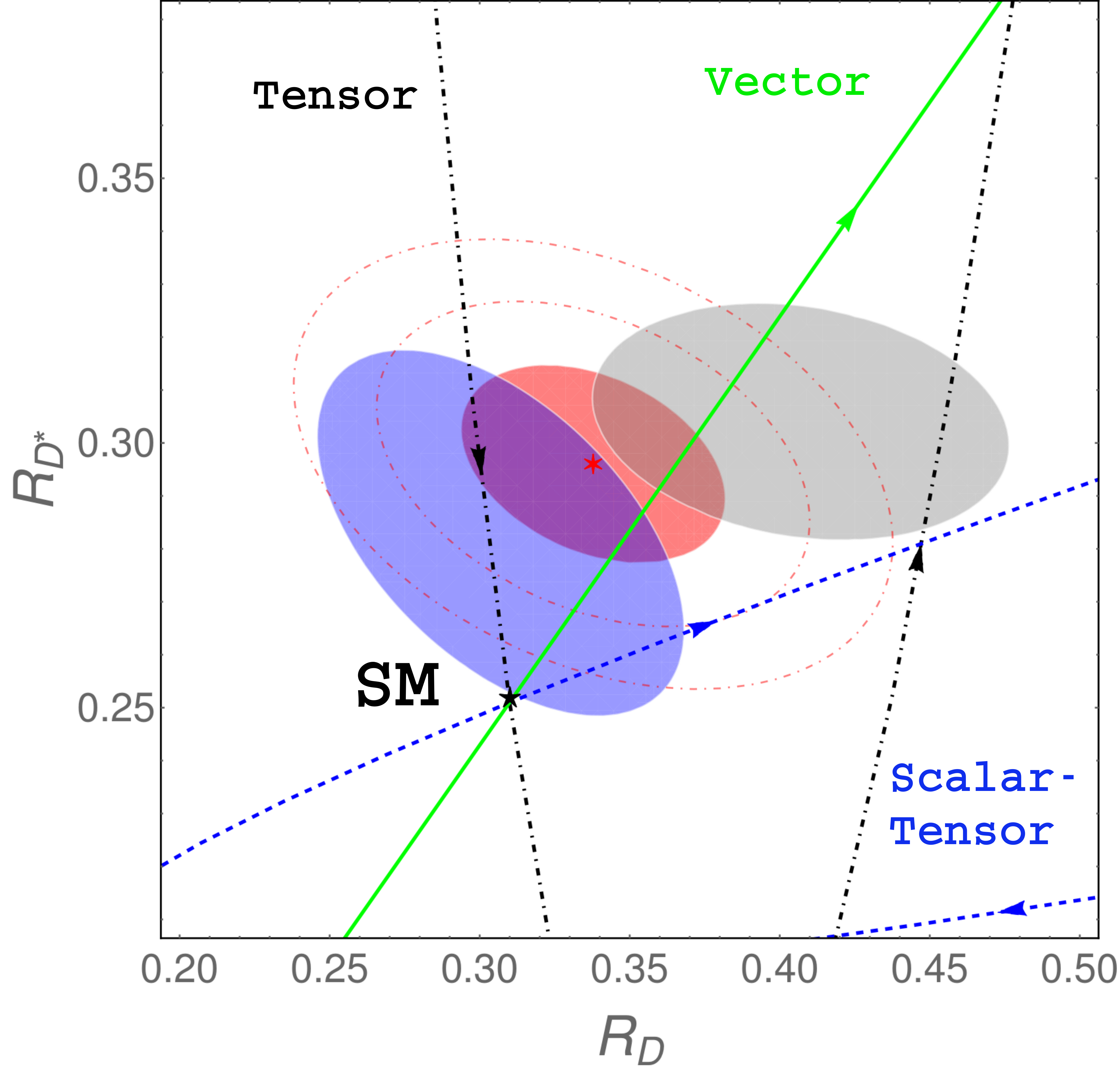}
\caption{Trajectories in the $(R_D,~R_{D^*})$ plane of predicted deviations from the SM due to NP where the arrows indicate the direction of positive increment of the WCs as defined in Eq.~\ref{eq:EFTLag}. ``Vector'' corresponds to either $\eL^\tau$ or $\tilde\eR^\tau$ while ``tensor'' and ``scalar-tensor'' correspond to $\eT^\tau$ and $\eSL^\tau=-4\eT^\tau$, respectively, at $\mu=1$ TeV and evolved down to $\mu=m_b$ using eq.~(\ref{eq:RGEepsilon}). The gray, blue and red solid ellipses are the 1$\sigma$ contours of the 2018 HFLAV average, the Belle measurement with semileptonic tag, and of the combination of the two, respectively. Red dot-dashed ellipses are $2\sigma$ and $3\sigma$ contours of the combination.}\label{fig:1}
\end{figure}

In Fig.~\ref{fig:1} we show the ``trajectories'' representing the correlated impact on $R_D$ and $R_{D^*}$ of NP scenarios where only a single operator
is present at a certain scale. Namely,  the ``vector'' curve is followed by scenarios with new pure left-handed ($\eL^\tau$) or pure right-handed ($\tilde \eR^\tau$) currents (which are not affected by short distance QCD corrections). ``Tensor'' and ``scalar-tensor'' interpretations involve both $\eT^\tau$ and $\eSL^\tau$ coupled by the radiative corrections in the SM, \textit{cf.} eq.~(\ref{eq:RGEepsilon}). The tensor trajectory describes a solution with only the tensor operator produced at the heavy scale (\textit{cf.} produced by the combination of $S_1$- and $R_2$-leptoquark contributions described in Sec.~\ref{sec:mediators}), that we take to be 1 TeV. The scalar-tensor description assumes the relation $\eSL^\tau(1\text{ TeV})=-4\eT^\tau(1\text{ TeV})$, again, at the heavy scale (\textit{cf.} produced by the $S_1$ leptoquark). The arrows in the curves signal the direction of positive increment of the WCs. The experimental data in Table~\ref{tab:1} is represented by the different ellipses: The gray one is the $1\sigma$ contour of the 2018 HFLAV average, the blue ellipse is the $1\sigma$ region of the 2019 Belle measurement with semi-leptonic tag and, finally, the red ellipses are the 1$\sigma$, 2$\sigma$ and 3$\sigma$ contours of the combination of these two.

The interference of the SM with left-handed or scalar-tensor contributions can produce a simultaneous increase of $R_{D}$ and $R_{D^*}$, as illustrated in Fig.~\ref{fig:1} by the positive slope of the corresponding curves at the SM point. This effect drives these solutions to agree well with the 2018 HFLAV average. In case of the tensor scenario, interference with the SM increases $R_{D}$ at the expense of reducing $R_{D^*}$ or vice versa. This effect is illustrated by the negative slope of the ``Tensor'' curve in Fig.~\ref{fig:1}. Therefore, the agreement of this scenario with the older data set is due to the quadratic contributions of the tensor operator to the rates. With the new Belle measurement, $R_{D}$ becomes more consistent with the SM while a value of $R_{D^*}$  larger than predicted is still favored. In this new scenario, ``vector'' models still agree with the data but now the interference of the tensor operator with the SM can play a role in providing a satisfactory solution.

\begin{table}[h!]
 \caption{\label{tab:fit1} Best fit values, $\chi_{\rm min}^2$, $p$-value, pull and $1\sigma$ confidence intervals of the WCs in the fits to all the $R_{D^{(*)}}$ data. We perform fits to one or two WCs at a time with the understanding that the others are set to 0.
 For the cases of two Wilson-coefficient fits, the 1$\sigma$ interval of each Wilson coefficient is obtained by profiling over the other one to take into account their correlation.\label{tab:3}}
\begin{center}
        \begin{tabular}{ccccccc}
      \hline
      \hline
      ~~~~~~~~ & ~~Best fit~~  & ~~$\chi_{\rm min}^2$~~  & ~~p-value~~ & ~~${\rm Pull}_{\rm SM}$~~ ~~ & $1 \sigma$ range~~\\
      \hline

      ~~$\epsilon_L^\tau$~~ & ~~$0.07$~~ & ~~$9.00$~~ & ~~$0.34$~~ & ~~$3.43$~~ & ~~$(0.05,0.09)$~~\\

      ~~$\epsilon_T^\tau$~~ & ~~$-0.03$~~ & ~~$9.85$~~ & ~~$0.28$~~ & ~~$3.30$~~ & ~~$(-0.04,-0.02)$~~\\

      ~~$\epsilon_{S_L}^\tau$~~ & ~~$0.09$~~ & ~~$19.14$~~ & ~~$1.41\times10^{-2}$~~ & ~~$1.27$~~ & ~~$(0.02,0.15)$~~\\

      ~~$\epsilon_{S_R}^\tau$~~ & ~~$0.13$~~ & ~~$15.84$~~ & ~~$4.47\times10^{-2}$~~ & ~~$2.22$~~ & ~~$(0.07,0.20)$~~\\

      ~~$\tilde\epsilon_R^\tau$~~ & ~~$0.38$~~ & ~~$9.00$~~ & ~~$0.34$~~ & ~~$3.43$~~ & ~~$(0.32,0.44)$~~\\

      ~~$\epsilon_{S_L}^\tau=-4\epsilon_T^\tau$~~ & ~~$0.09$~~ & ~~$12.25$~~ & ~~$0.14$~~ & ~~$2.92$~~ & ~~$(0.06,0.12)$~~\\

      ~~$(\epsilon_{S_L}^\tau,\epsilon_T^\tau)$~~ & ~~$(0.07,-0.03)$~~ & ~~$8.7$~~ & ~~$0.27$~~ & ~~$3.03$~~ & ~~$\epsilon_{S_L}^\tau\in(0.00,0.14)$~~ ~~$\epsilon_T^\tau\in(-0.04,-0.02)$~~\\

      ~~$(\epsilon_{S_L}^\tau,\epsilon_{S_R}^\tau)$~~ & ~~$(-0.47,0.53)$~~ & ~~$8.7$~~ & ~~$0.27$~~ & ~~$3.03$~~ & ~~$\epsilon_{S_L}^\tau\in(-0.66,-0.30)$~~ ~~$\epsilon_{S_R}^\tau\in(0.37,0.69)$~~\\

      ~~$(\epsilon_{S_R}^\tau,\epsilon_T^\tau)$~~ & ~~$(0.07,-0.03)$~~ & ~~$8.7$~~ & ~~$0.27$~~ & ~~$3.03$~~ & ~~$\epsilon_{S_R}^\tau\in(0.00,0.14)$~~~~$\epsilon_T^\tau\in(-0.04,-0.02)$~~\\

      ~~$(\epsilon_L^\tau,\epsilon_T^\tau)$~~ & ~~$(0.05,-0.01)$~~ & ~~$8.7$~~ & ~~$0.27$~~ & ~~$3.03$~~ & ~~$\epsilon_L^\tau\in(0.00,0.09)$~~~~$\epsilon_T^\tau\in(-0.03,0.01)$~~\\

      ~~$(\epsilon_L^\tau,\epsilon_{S_L}^\tau)$~~ & ~~$(0.08,-0.04)$~~ & ~~$8.7$~~ & ~~$0.27$~~ & ~~$3.03$~~ & ~~$\epsilon_L^\tau\in(0.05,0.10)$~~~~$\epsilon_{S_L}^\tau\in(-0.13,0.04)$~~\\

      ~~$(\epsilon_L^\tau,\epsilon_{S_R}^\tau)$~~ & ~~$(0.08,-0.05)$~~ & ~~$8.7$~~ & ~~$0.27$~~ & ~~$3.03$~~ & ~~$\epsilon_L^\tau\in(0.05,0.11)$~~~~$\epsilon_{S_R}^\tau\in(-0.15,0.04)$~~\\
      \hline
      \hline
    \end{tabular}
  \end{center}
\end{table}

\begin{figure}[h!]
\centering
\includegraphics[width=12cm]{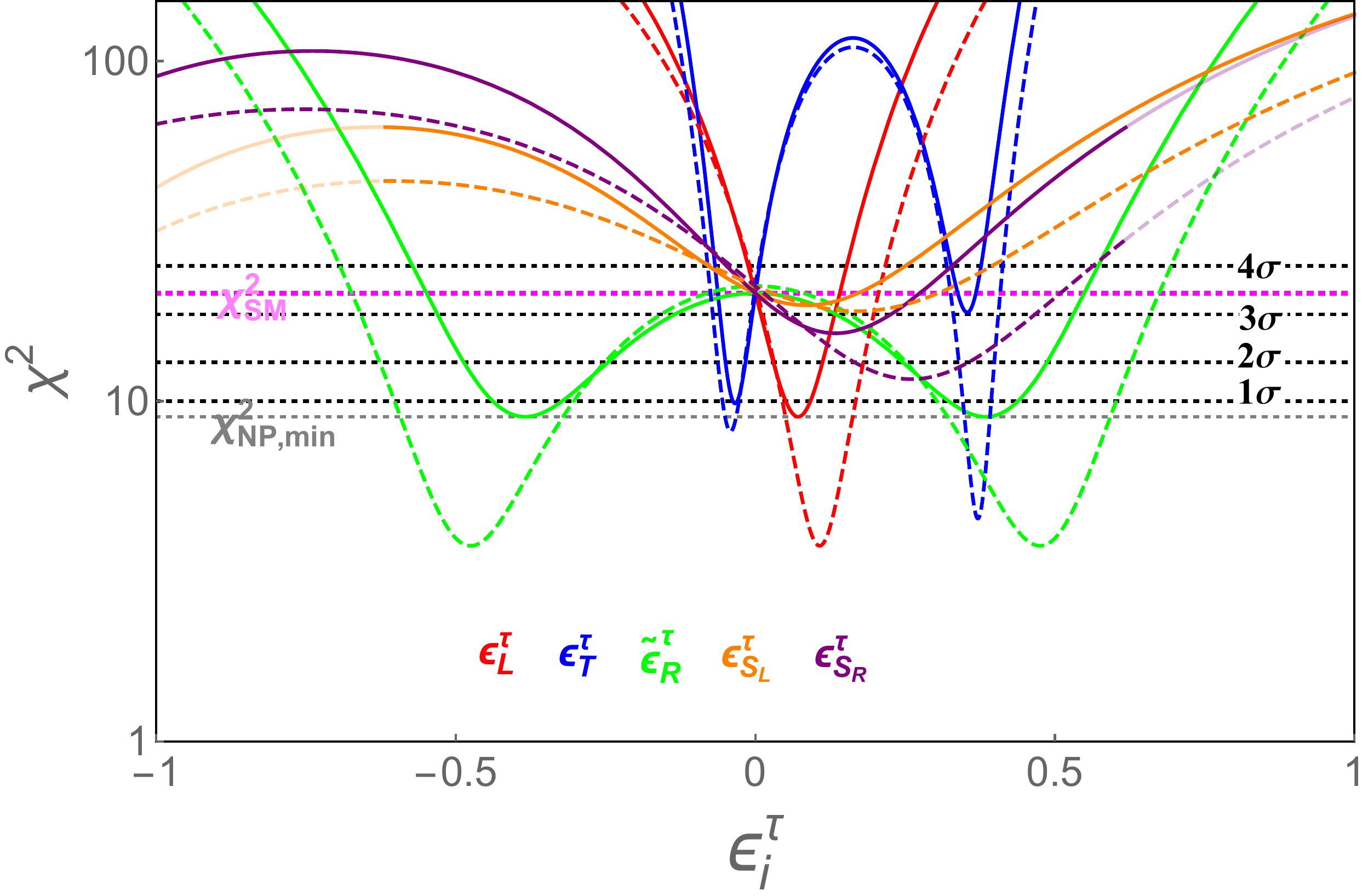}
\caption{The $\chi^2$ of the fits to $R_D$ and $R_{D^*}$ with one Wilson coefficient active at a time (setting the others to 0) and evaluated at the renormalization scale $\mu=m_b$. The solid lines correspond to the fits to the 2019 HFLAV average. Horizontal lines show the value at the minima of the model giving the best fit (vector scenario) and the 1$\sigma$ to 4$\sigma$ ranges computed from there. We also show the line corresponding to the value of $\chi^2_{\rm SM}$. The dashed lines correspond to the fits to the 2018 HFLAV average. Faded regions for $\epsilon_{S_L}^\tau$ and $\epsilon_{S_R}^\tau$ represent a exclusion of 30\% limit on ${\rm Br}(B_c\to\tau\nu)$.
}\label{fig:2}
\end{figure}

In Table~\ref{tab:fit1} we show the results of fits to all the data on $R_{D^{(*)}}$ of one or two WCs at a time, while setting the others to zero. In the two-dimensional case we only investigate the interplay between operators with left-handed neutrinos. Setting all WCs to zero, one obtains a $\chi_{\rm SM}^2=20.75$. With 9 degrees of freedom (d.o.f) this corresponds to a $p$-value of $1.38\times10^{-2}$. As can be inferred from the table, the ``vector'' operators provide the best one-parameter fit to the data, with a $p$-value of 0.34 and a SM pull of 3.43$\sigma$. The difference in size of the values of the WCs between the left- and right-handed vector solutions is due to the fact that the latter corresponds to a quadratic NP effect in the rates.

The tensor operator also gives a good fit to the data, where the solution driven by the interference piece is now preferred. Scalar models do not provide good fits and require values that may be in conflict with the bounds from $B_c\to \tau\nu$. In Fig.~\ref{fig:2} we show the functions $\chi^2$ of the one-parameter fits for each of the WCs. We also show in dashed lines the results obtained from the fits to the 2018 HFLAV average, to emphasize the change in the structure and values of the WCs needed with the new data. Horizontal lines showing the values of the 1- to 4$\sigma$ ranges have been computed taking the best model (vector operators) giving $\chi^2_{\rm min}=9.00$ as reference.

\begin{figure}[h!]
  \centering
  \includegraphics[width=16cm]{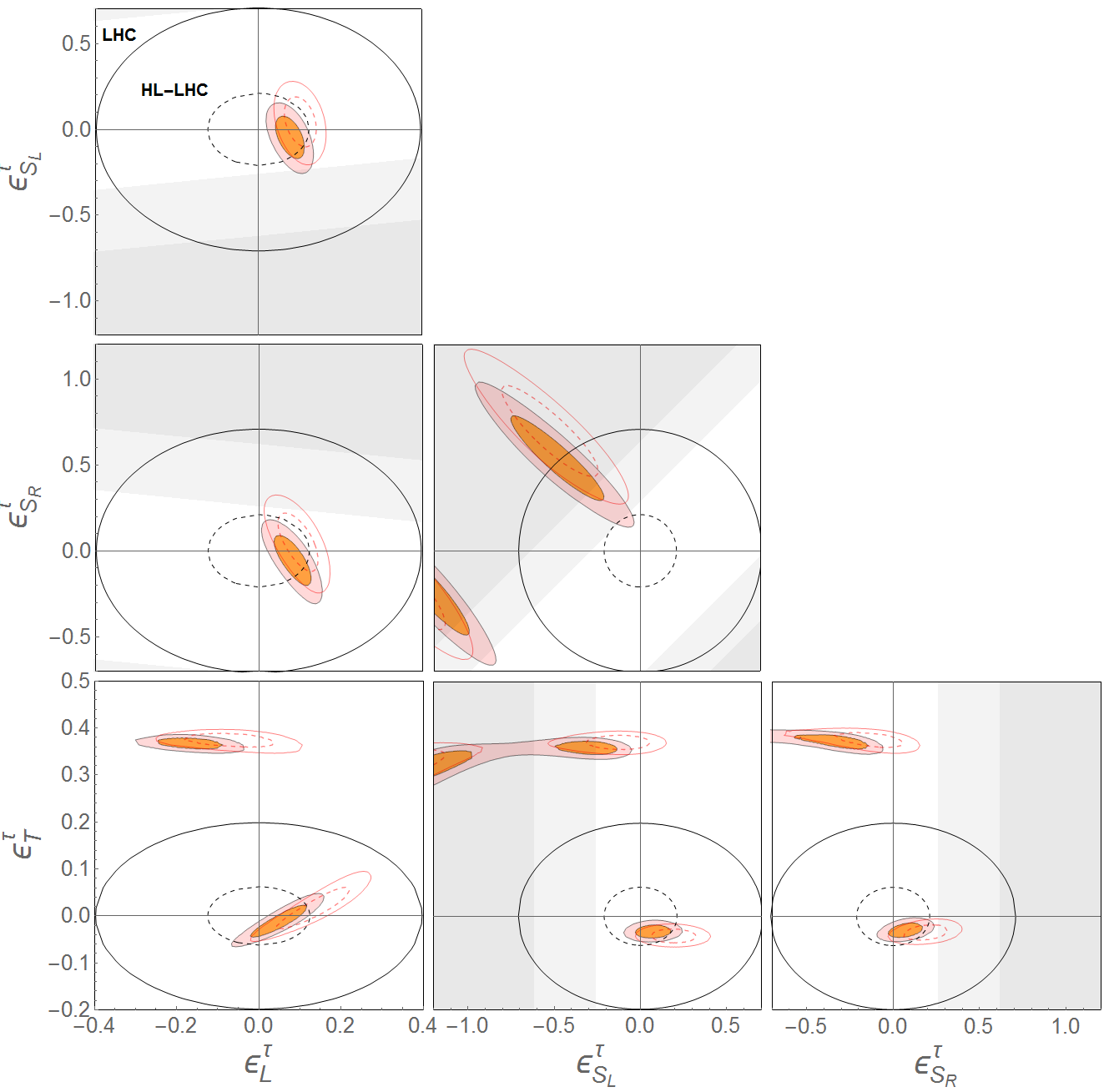}
  \caption{Constraints from the fits to $R_D$ and $R_{D^*}$ with two WCs active at a time (setting the others to 0) and evaluated at the renormalization scale $\mu=m_b$. Solid ellipses (empty red ellipses) represent 1$\sigma$ and 2$\sigma$ allowed regions from the fits to all the data (2018 HFLAV average). The empty black solid (dashed) ellipses indicate the 2$\sigma$ upper bounds from the LHC data (HL-LHC projections) on $pp\to\tau_h X$+MET. Regions in gray and light gray represent 30\% and 10\% exclusion limits from ${\rm Br}(B_c\to\tau\nu)$, respectively.}\label{fig:3}
\end{figure}

In Fig.~\ref{fig:3} we show the contour plots that are obtained from each of the six two-dimensional fits to  the 2019 HFLAV averages of $R_D$ and $R_{D^*}$. In the Appendix, Table~\ref{tab:fit1corr}, we provide the correlation matrices for the fits to two WCs. We also show with empty red contours the results of the fits to the 2018 HFLAV averages. Black empty contours represent the 2$\sigma$ upper limits that can be set by analyzing the tails of $p p\to \tau X$+MET at the LHC (solid line) and by estimating the projected sensitivity at the HL-LHC (dashed line)~\cite{Greljo:2018tzh}.

Adding the new Belle data in the fit results in regions which are slightly closer to the SM, although all NP scenarios still describe the data better with a significance of $3.03\sigma$. As expected, constraints from ${\rm Br}(B_c\to\tau\nu)$ play an important role in excluding regions of the parameter space of the scalar models. For instance, in case of the pure scalar fit, with $(\eSL^\tau,~\eSR^\tau)$, the $1\sigma$ region is almost excluded by the softer limit based on the $B_c$ lifetime. Even the $2\sigma$ region is also excluded if the more aggressive limit of 10\%  on ${\rm Br}(B_c\to\tau\nu)$ is used. Constraints in the $(\eSL^\tau,~\eT^\tau)$ plane are interesting for UV completions involving $S_1$ and $R_2$ leptoquarks. In this scenario, data favors the parameter space in which the two WCs have the opposite  sign, like the contribution of the $S_1$ and unlike the one of the  $R_2$, \textit{cf.} eqs.~(\ref{eq:LagEFTS1}) and eq.~(\ref{eq:LagEFTR2}). A fit with the scalar-tensor contribution produced by the $S_1$ leptoquark (evaluated at $\mu=1$ TeV) gives a fit with a $p$-value 0.15 that is considerably better than for the SM. However, this scenario performs worse than those with pure left-handed or tensor operators. Constraints in the $(\eL^\tau,~\eSR^\tau)$ plane are interesting for UV completions of the $U_1$ leptoquark involving left- and right-handed currents to the fermions~\cite{Bordone:2017bld,Bordone:2018nbg,Baker:2019sli,Cornella:2019hct}.

The LHC data also probes the parameter space of the preferred regions in the different scenarios. As already anticipated in~\cite{Greljo:2018tzh}, scenarios involving large quadratic contributions of the tensor operator are excluded by more than 2$\sigma$. Furthermore, the current LHC exclusion region independently covers a large portion  of the 1$\sigma$ ellipse in the pure scalar scenario and all the parameter space of the $2\sigma$ region will be probed by the HL-LHC. In fact, with the high-luminosity data set we should be able to probe all the interesting regions in all the scenarios, although less deeply than for the results of the fits to the 2018 HFLAV average.

A potential caveat concerning the interpretation of these LHC bounds is that their validity relies on the assumption that the NP scale is significantly larger than the partonic energies probing the effective interaction in the $p p \to \tau \nu$ collisions at the LHC. In ref.~\cite{Greljo:2018tzh} this was studied by assessing the sensitivity to NP of the distribution in the tau transverse-mass, $m_T$, of the $p p \to \tau_h X$+MET analyses~\cite{Aaboud:2018vgh,Sirunyan:2018lbg}. Most of the sensitivity of the LHC stems from $m_T\lesssim2$ TeV and, for mediator masses above this mark, the EFT provides a faithful description of the NP signal. By taking the central values of the one-parameter fits shown in Tab.~\ref{tab:fit1}, and assuming $\mathcal O(1)$ couplings in eqs.~(\ref{eq:LagEFTWp}), (\ref{eq:LagEFTS1}), (\ref{eq:LagEFTR2}) and (\ref{eq:LagEFTU1}) we find that the masses of the putative new mediators are $m_{S_1}\simeq 2.3$ TeV, $m_{U_1}\simeq3.3$ TeV for left-handed current couplings and approximately a factor two lighter for right-handed current couplings, {\it cf.} $m_{W^\prime}\simeq1.4$ TeV. For the tensor scenario, $m_{S_1}\simeq m_{R_2}\simeq2.3$ TeV. Therefore, in the comparison with the LHC bounds shown Fig.~\ref{tab:fit1} we are implicitly assuming that the mediators are in this regime of couplings and masses.

Extending the comparison to the right-handed currents, the value $\tilde \epsilon_R^\tau=0.38(6)$ obtained in the fit would still be challenged by the bound $|\tilde \epsilon_R^\tau|\leq0.32$ at $2\sigma$ resulting from the collider analysis in the EFT. Turning to explicit UV completions in the range of masses below 2 TeV, LHC bounds are stronger than the EFT counterpart for the $W^\prime$ but weaker for the leptoquarks~\cite{Greljo:2018tzh}~\footnote{For a reanalysis of the impact of the 2019 Belle data in the collider bounds using the monotau searches in the models addressing the $R_{D^{(*)}}$ anomalies see~\cite{JMCPortoroz2019}.}
\begin{figure}[h]
\begin{tabular}{ccc}
  \includegraphics[width=70mm]{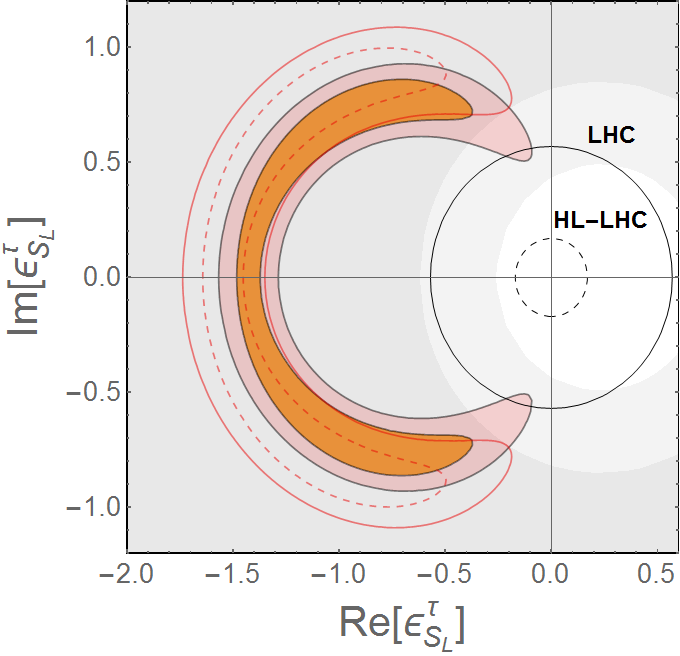}
& \hspace{0.8cm} &
 \includegraphics[width=70mm]{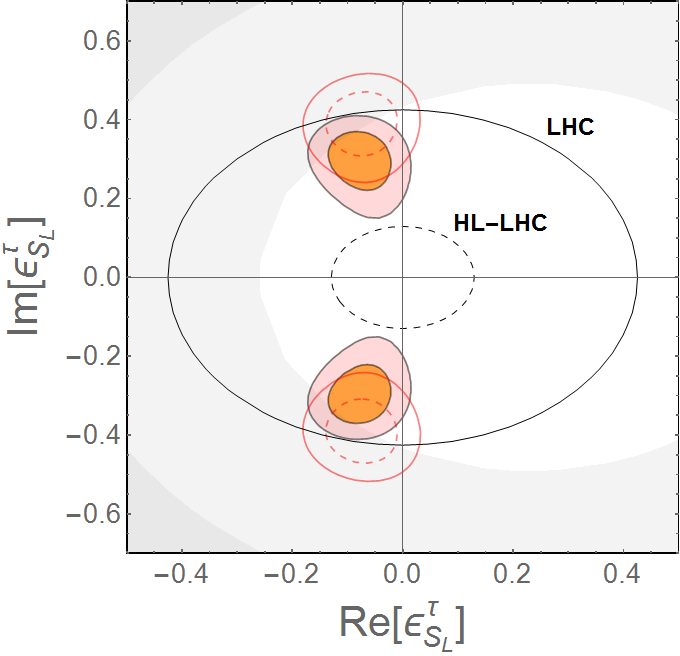}
\end{tabular}
\caption{Constraints from the fits to $R_D$ and $R_{D^*}$ on the complex $\epsilon^\tau_{S_L}$ plane evaluated at the renormalization scale $\mu=m_b$: \textit{Left:} All WCs other than $\epsilon^\tau_{S_L}$  are set to 0; \textit{Right:} The condition $\epsilon^\tau_{S_L}=4 \epsilon^\tau_{T}$ is imposed at the matching scale $\mu=M_{R_2}=$1~TeV, as in the $R_2$-mediator model, and all  WCs other than $\epsilon^\tau_{S_L}$ and $\epsilon^\tau_{T}$  are set to 0.  Solid regions (empty red regions) represent 1$\sigma$ and 2$\sigma$ allowed regions from the fits to the 2019 (2018) HFLAV average of $R_{D^{(*)}}$ data. The empty black solid (dashed) ellipses indicate the 2$\sigma$ upper bounds from the LHC data (HL-LHC projections) on $pp\to\tau_h X$+MET. Regions in gray and light gray represent 30\% and 10\% exclusion limits from ${\rm Br}(B_c\to\tau\nu)$, respectively.}
\label{fig:complexWCs}
\end{figure}

Finally, in our fits we have assumed real WCs, but is clear from Sec.~\ref{sec:mediators} that these are generally complex. The imaginary part of WCs does not interfere with the leading, SM contribution. Hence it is expected that fits with  purely imaginary WCs are susceptible to bounds from, eg, the rate  $B_c\to\tau\nu$. This is particularly the case for scalar WCs, that require large magnitude of WCs to account for $R_{D^{(*)}}$. For example, in the left panel of Fig.~\ref{fig:complexWCs} we show the fit of the complex WC $\epsilon_{S_L}$ with best fit value located at $\epsilon_{S_L}=-0.88\pm0.74~i$ and where the $2\sigma$ C.L. region is excluded by the $B_c$ lifetime and LHC constraints. However, in some cases allowing a complex phase may improve a WC fit. An interesting example is that of the combination $\epsilon^\tau_{S_L}=4 \epsilon^\tau_{T}$ that is the case of the $R_2$ leptoquark mediator, Eq.~(\ref{eq:LagEFTR2}); the right panel of Fig.~\ref{fig:complexWCs} shows the constraints on the complex $\epsilon^\tau_{S_L}(\mu=m_b)$ plane, with best fit point at $\epsilon_{S_L}=-0.08\pm0.30~i$, and having imposed the condition $\epsilon^\tau_{S_L}=4 \epsilon^\tau_{T}$ at the matching scale $\mu=M_{R_2}=1~\text{TeV}$.\footnote{Complex coefficients for a model based on $R_2$ were considered in Ref.~\cite{Becirevic:2018afm}. }

\subsection{Fits to $R_D$, $R_{D^*}$, $R_{J/\psi}$, $P_\tau^{D^*}$ and $F_L^{D^*}$ data}
\label{sec:globalfits}

In this section, we perform a global fit of $\epsilon_L^\tau$, $\epsilon_T^\tau$, $\epsilon_{S_L}^\tau$ and $\epsilon_{S_R}^\tau$ to all the data including $R_D$ and $R_{D^*}$, $R_{J/\psi}$, $P_\tau^{D^*}$ and $F_L^{D^*}$. We implement the LHC monotau constraints by demanding that the WCs are within the corresponding 2$\sigma$ bounds, i.e., we take $|\epsilon_L^\tau|\leq0.32$, $|\epsilon_T^\tau|\leq0.16$, $|\epsilon_{S_L}^\tau|\leq0.57$ and $|\epsilon_{S_R}^\tau|\leq0.57$. In addition, we impose the constraint from the $B_c$ lifetime by requiring that ${\rm Br}(B_c\to\tau\nu)\leq30\%$. One obtains a $\chi_{\rm min,SM}^2=26.53$ with 12 degrees of freedom (d.o.f) if all the WCs are set to 0, corresponding to a $p$-value of $9.02\times10^{-3}$. The resulting WCs from the fit are,
\begin{align}\label{eq:globalfit_results}
 \left(
\begin{array}{c}
\eL^\tau\\
\eT^\tau\\
\eSL^\tau\\
\eSR^\tau
\end{array}
\right)=\left(
\begin{array}{c}
0.16\pm0.20\\
0.05\pm0.09\\
-0.33\pm0.21\\
0.14\pm0.22
\end{array}
\right),
\end{align}
with the correlation matrix,
\begin{align}\label{eq:globalfit_rho}
 \rho=\left(
\begin{array}{cccc}
 1.000 & 0.816 & 0.913 & -0.915\\
  & 1.000 & 0.951& -0.920\\
  & & 1.000& -0.986\\
 &  & & 1.000
\end{array}
\right),
\end{align}
and where $\chi_{\rm min}^2=12.80$ for 8 d.o.f., corresponding to a $p$-value of 0.12 and a Pull$_{\rm SM}=2.64$. This provides an approximation of the $\chi_{\rm min}^2$ in the immediate vicinity of the minimum that is closest to the SM, although it is not appropriate to obtain realistic confidence-level regions. For instance, the $1\sigma$ intervals may seem to violate the LHC bounds described above.
\begin{figure}[h!]
\centering
\includegraphics[width=16cm]{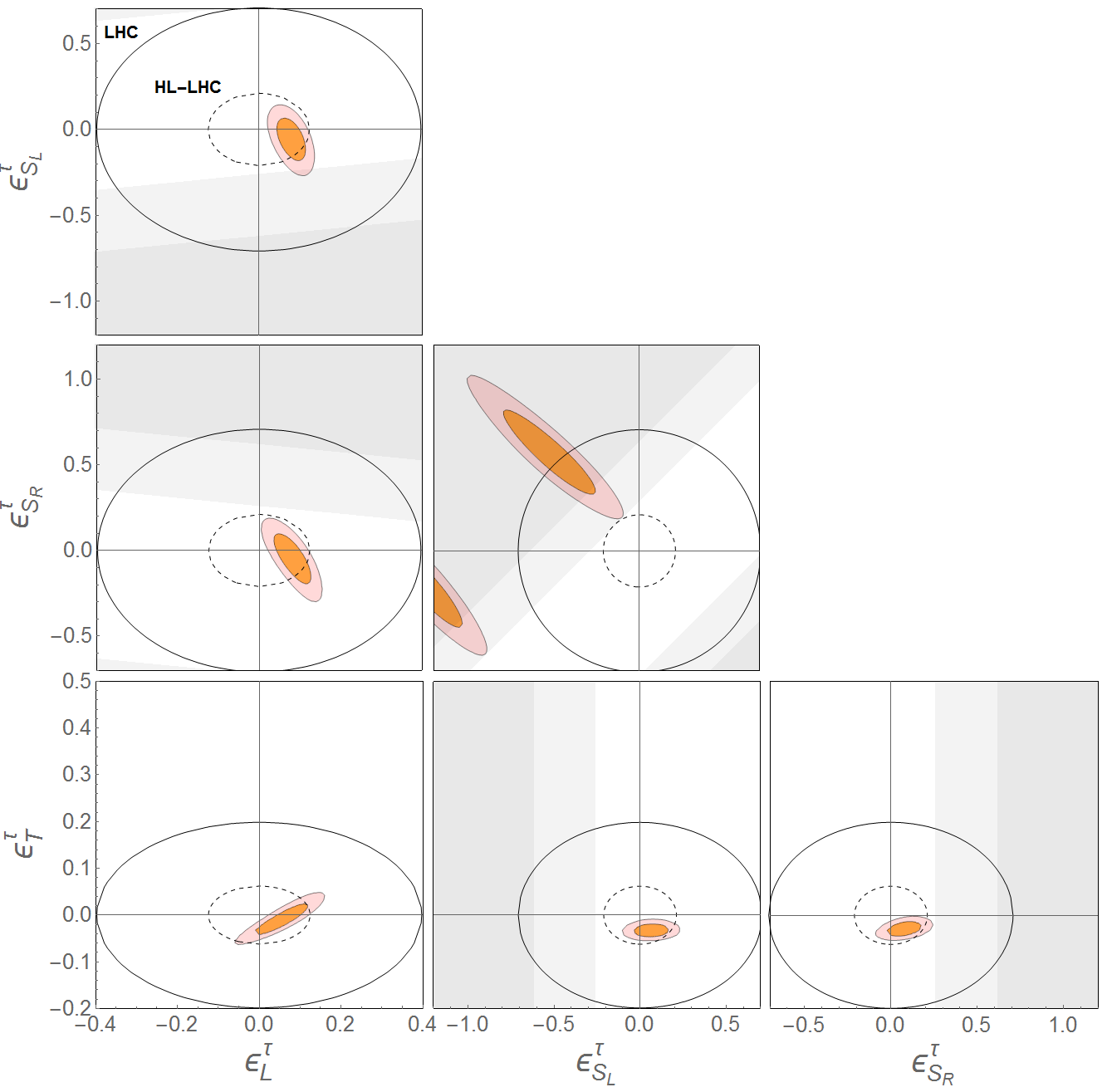}
\caption{Constraints in the WCs planes from the fits to all the data in $R_D$ and $R_{D^*}$, and to $R_{J/\psi}$, $P_\tau^{D^*}$ and $F_L^{D^*}$ setting two WCs to zero. The solid ellipses represent 1$\sigma$ and 2$\sigma$ allowed regions while the empty black solid (dashed) ellipses indicate the 2$\sigma$ upper bounds from the LHC data (HL-LHC projections) on $pp\to\tau_h X$+MET. Regions in gray and light gray represent 30\% and 10\% exclusion limits from ${\rm Br}(B_c\to\tau\nu)$, respectively.}\label{fig:4}
\end{figure}
\begin{table}[h!]
 \caption{\label{tab:fit2} Best fit values, $\chi_{\rm min}^2$, $p$-value, pull and $1\sigma$ confidence intervals of the WCs in the fits to all the data in $R_D$, $R_{D^*}$,  $R_{J/\psi}$, $P_\tau^{D^*}$ and $F_L^{D^*}$. We perform fits to one or two WCs at a time with the understanding that the others are set to 0. For the cases of two WC fits, to take into account correlation between the two WCs, the 1$\sigma$ interval of each WC  is obtained by profiling over the other WC.}
\begin{center}
     \begin{tabular}{cccccc}
      \hline
      \hline
      ~~~~~~~~ & ~~Best fit~~  & ~~$\chi_{\rm min}^2$~~  & ~~p-value~~ & ~~${\rm Pull}_{\rm SM}$~~ & ~~1$\sigma$ range~~\\
      \hline

      ~~$\epsilon_L^\tau$~~ & ~~$0.07$~~ & ~~$14.56$~~ & ~~$0.20$~~ & ~~$3.46$~~ & ~~$(0.05,0.09)$~~\\

      ~~$\epsilon_T^\tau$~~ & ~~$-0.03$~~ & ~~$15.70$~~ & ~~$0.15$~~ & ~~$3.29$~~ & ~~$(-0.04,-0.02)$~~\\

      ~~$\epsilon_{S_L}^\tau$~~ & ~~$0.08$~~ & ~~$25.23$~~ & ~~$8.44\times10^{-3}$~~ & ~~$1.14$~~ & ~~$(0.01,0.14)$~~\\

      ~~$\epsilon_{S_R}^\tau$~~ & ~~$0.14$~~ & ~~$21.24$~~ & ~~$3.10\times10^{-2}$~~ & ~~$2.30$~~ & ~~$(0.08,0.20)$~~\\

      ~~$(\epsilon_{S_L}^\tau,\epsilon_T^\tau)$~~ & ~~$(0.07,-0.03)$~~ & ~~$14.75$~~ & ~~$0.14$~~ & ~~$3.00$~~ & ~~$\epsilon_{S_L}^\tau\in(0.00,0.13)$~~~~$\epsilon_T^\tau\in(-0.04,-0.02)$~~\\

      ~~$(\epsilon_{S_L}^\tau,\epsilon_{S_R}^\tau)$~~ & ~~$(-0.51,0.56)$~~ & ~~$12.14$~~ & ~~$0.28$~~ & ~~$3.37$~~ & ~~$\epsilon_{S_L}^\tau\in(-0.69,-0.34)$~~~~$\epsilon_{S_R}^\tau\in(0.41,0.73)$~~\\

      ~~$(\epsilon_{S_R}^\tau,\epsilon_T^\tau)$~~ & ~~$(0.08,-0.03)$~~ & ~~$14.38$~~ & ~~$0.16$~~ & ~~$3.05$~~ & ~~$\epsilon_{S_R}^\tau\in(0.01,0.14)$~~~~$\epsilon_T^\tau\in(-0.04,-0.02)$~~\\

      ~~$(\epsilon_L^\tau,\epsilon_T^\tau)$~~ & ~~$(0.05,-0.01)$~~ & ~~$14.32$~~ & ~~$0.16$~~ & ~~$3.06$~~ & ~~$\epsilon_L^\tau\in(0.01,0.10)$~~ ~~$\epsilon_T^\tau\in(-0.03,0.01)$~~\\

      ~~$(\epsilon_L^\tau,\epsilon_{S_L}^\tau)$~~ & ~~$(0.08,-0.06)$~~ & ~~$14.09$~~ & ~~$0.17$~~ & ~~$3.09$~~ & ~~$\epsilon_L^\tau\in(0.06,0.10)$~~ ~~$\epsilon_{S_L}^\tau\in(-0.14,0.03)$~~\\

      ~~$(\epsilon_L^\tau,\epsilon_{S_R}^\tau)$~~ & ~~$(0.08,-0.05)$~~ & ~~$14.33$~~ & ~~$0.16$~~ & ~~$3.06$~~ & ~~$\epsilon_L^\tau\in(0.05,0.11)$~~ ~~$\epsilon_{S_R}^\tau\in(-0.14,0.05)$~~\\
      \hline
      \hline
    \end{tabular}
  \end{center}
  \end{table}

\begin{table}[h!]
 \caption{\label{tab:fit3}Different confidence-level intervals of the WCs in the fits to all the data in $R_D$, $R_{D^*}$,  $R_{J/\psi}$, $P_\tau^{D^*}$ and $F_L^{D^*}$, obtained from the profile $\chi^2$ where the rest of WCs are minimized within the $2\sigma$ LHC mono-tau bound. We have also applied the 30\% bound on ${\rm Br}(B_c\to\tau\nu)$.}
\begin{center}
    \begin{tabular}{ccccc}
      \hline
      \hline
      ~~~~~~~~ & ~~Best fit~~ & ~~1$\sigma$ range~~& ~~2$\sigma$ range~~& ~~3$\sigma$ range~~\\
      \hline

      ~~$\epsilon_L^\tau$~~ & ~~$0.16$~~ & ~~$(-0.04,0.36)$~~ & ~~$(-0.41,0.42)$~~ & ~~$(-0.45,0.47)$\\

      ~~$\epsilon_T^\tau$~~ & ~~$0.05$~~ & ~~$(-0.04,0.14)$~~ & ~~$(-0.14,0.18)$~~ & ~~$(-0.15,0.23)$~~\\

      ~~$\epsilon_{S_L}^\tau$~~ & ~~$-0.33$~~ & ~~$(-0.54,-0.12)$~~ & ~~$(-1.07,0.57)$~~ & ~~$(-1.11,0.76)$~~\\

      ~~$\epsilon_{S_R}^\tau$~~ & ~~$0.14$~~ & ~~$(-0.08,0.36)$~~ & ~~$(-1.27,0.57)$~~ & ~~$(-1.34,0.62)$~~\\
      \hline
      \hline
    \end{tabular}
  \end{center}

\end{table}

\begin{figure}[h!]
\centering
\includegraphics[width=16cm]{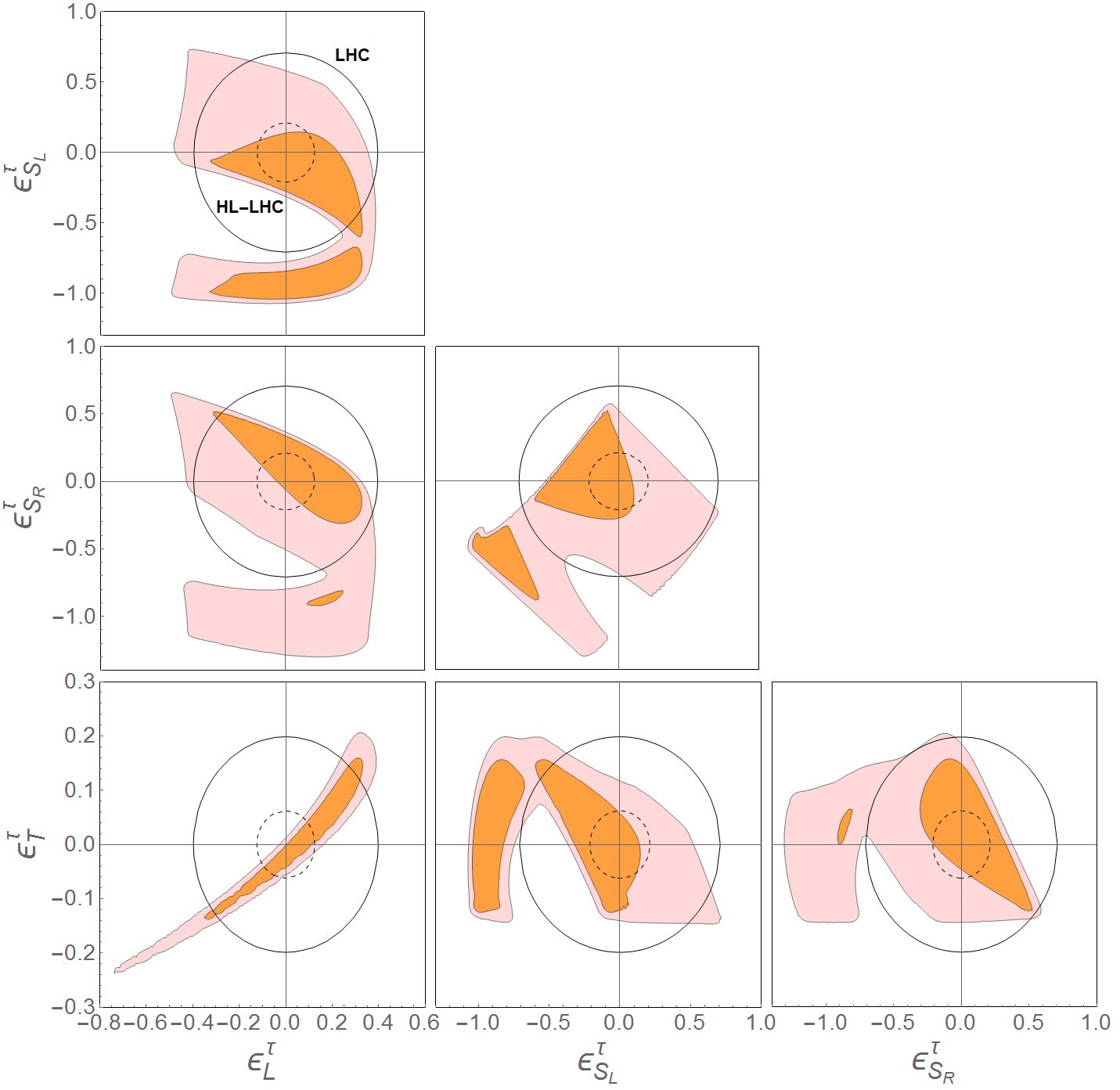}
\caption{Constraints in the WCs planes from the fits to all the data in $R_D$, $R_{D^*}$,  $R_{J/\psi}$, $P_\tau^{D^*}$ and $F_L^{D^*}$ profiling over the other WCs. The solid ellipses represent 1$\sigma$ and 2$\sigma$ allowed regions while the empty black solid (dashed) ellipses indicate the 2$\sigma$ upper bounds from the LHC data (HL-LHC projections) on $pp\to\tau_h X$+MET. Note that we have considered the  30\% bound on ${\rm Br}(B_c\to\tau\nu)$.}\label{fig:5}
\end{figure}

In order to investigate this in more detail we perform, first, fits of two WCs to $R_D$ and $R_{D^*}$, $R_{J/\psi}$, $P_\tau^{D^*}$ and $F_L^{D^*}$ setting the others to 0. This allows one to compare to the results of the two-parameters fits to $R_D$ and $R_{D^*}$ presented in Sec III-A. The corresponding six possible combinations of two WCs fits are shown in Fig.~\ref{fig:4} and the results of the fits are shown in Table~\ref{tab:fit2}. In the Appendix, Table~\ref{tab:fit2corr}, we provide the correlation matrices for these fits. As compared with Fig.~\ref{fig:3}, one notes that although not precise, the data $R_{J/\psi}$, $P_\tau^{D^*}$ and $F_L^{D^*}$ is sensitive enough to exclude the same regions allowed at $2\sigma$ by the fit to $R_{D^{(*)}}$ independently excluded by the LHC monotau signature or $B_c\to \tau\nu$ (see also Ref.~\cite{Aebischer:2018iyb}). However, for the favored regions of the fits closer to the SM the addition of the current data on these observables has a small impact.

Finally, in order to obtain realistic confidence-level regions with the four active WCs we obtain profile likelihoods functions depending on one or two WCs at a time. The monotau LHC constraints and the $B_c$ lifetime bound are implicitly imposed when profiling over the other "nuisance" WCs in each case. In Fig.~\ref{fig:5}, we show the results of the fits as constraints in the six two-WCs plots. In Tab.~\ref{tab:fit3} we show the final $1\sigma$, $2\sigma$ and $3\sigma$ confidence-level intervals for the WCs. The $1\sigma$ intervals are consistent with those obtained from
the fit in eq.~\ref{eq:globalfit_results}, while the $2\sigma$ and $3\sigma$ intervals differ from those obtained using the gaussian approximation of the $\chi^2$.

\subsection{The sensitivity of observables to New Physics}
\label{sec:sensitivity}

As shown above, different NP scenarios currently give a good description of the data, so the natural question is which other observables, beyond $R_{D}$ and $R_{D^*}$, allow one to discriminate among them. Only total rates are sensitive to the effects from the vector operators as their effects cancel in normalized observables. On the other hand, scalar and tensor operators change the kinematic distributions of the decays and show up in observables such as tau and recoiling-hadron polarizations (if the latter carries spin), $q^2$-distribution of the rate or angular analyses.

%

\begin{figure}[htpb]
  \centering
  \includegraphics[width=5cm,height=3.5cm]{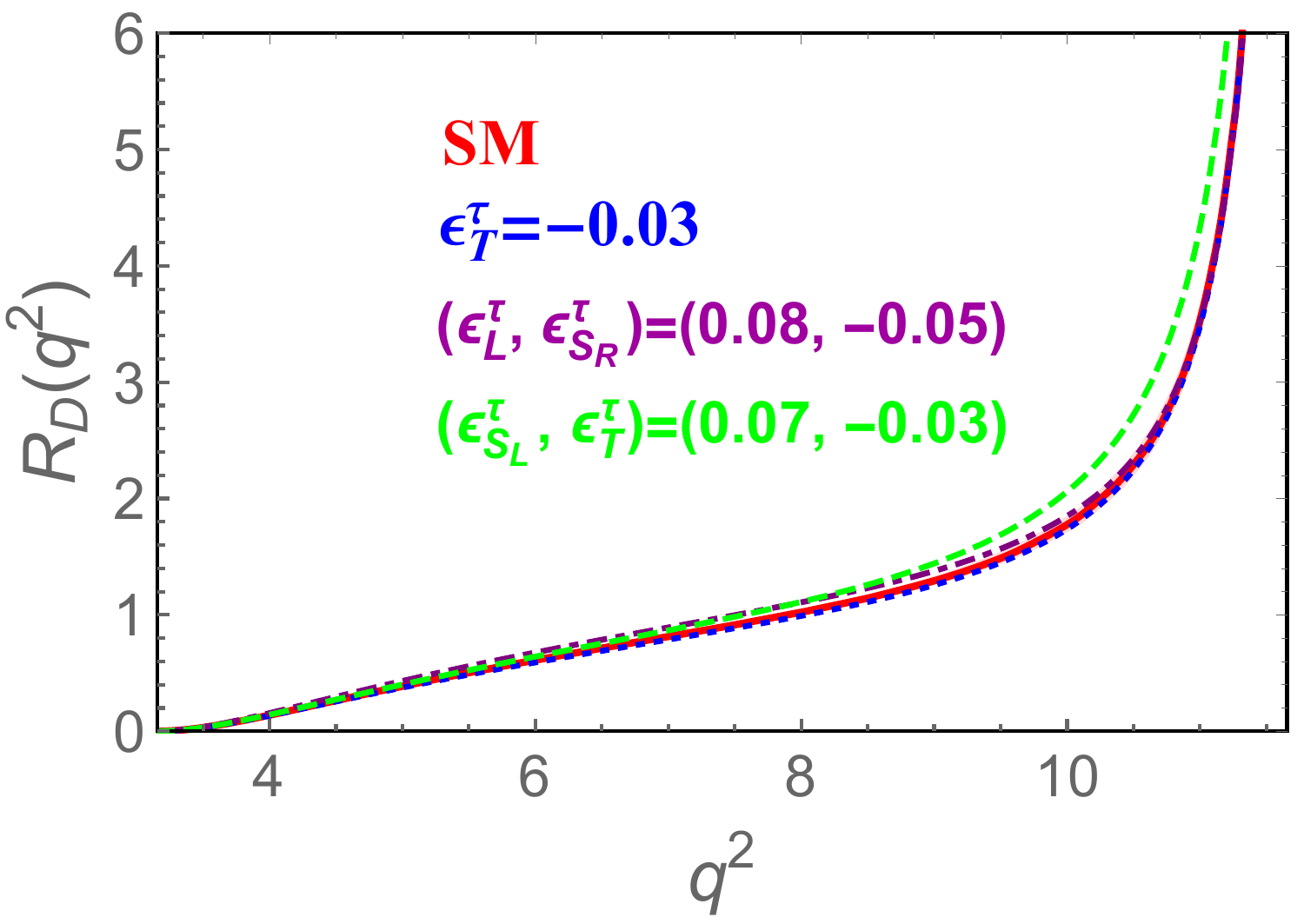}~~~\includegraphics[width=5cm,height=3.5cm]{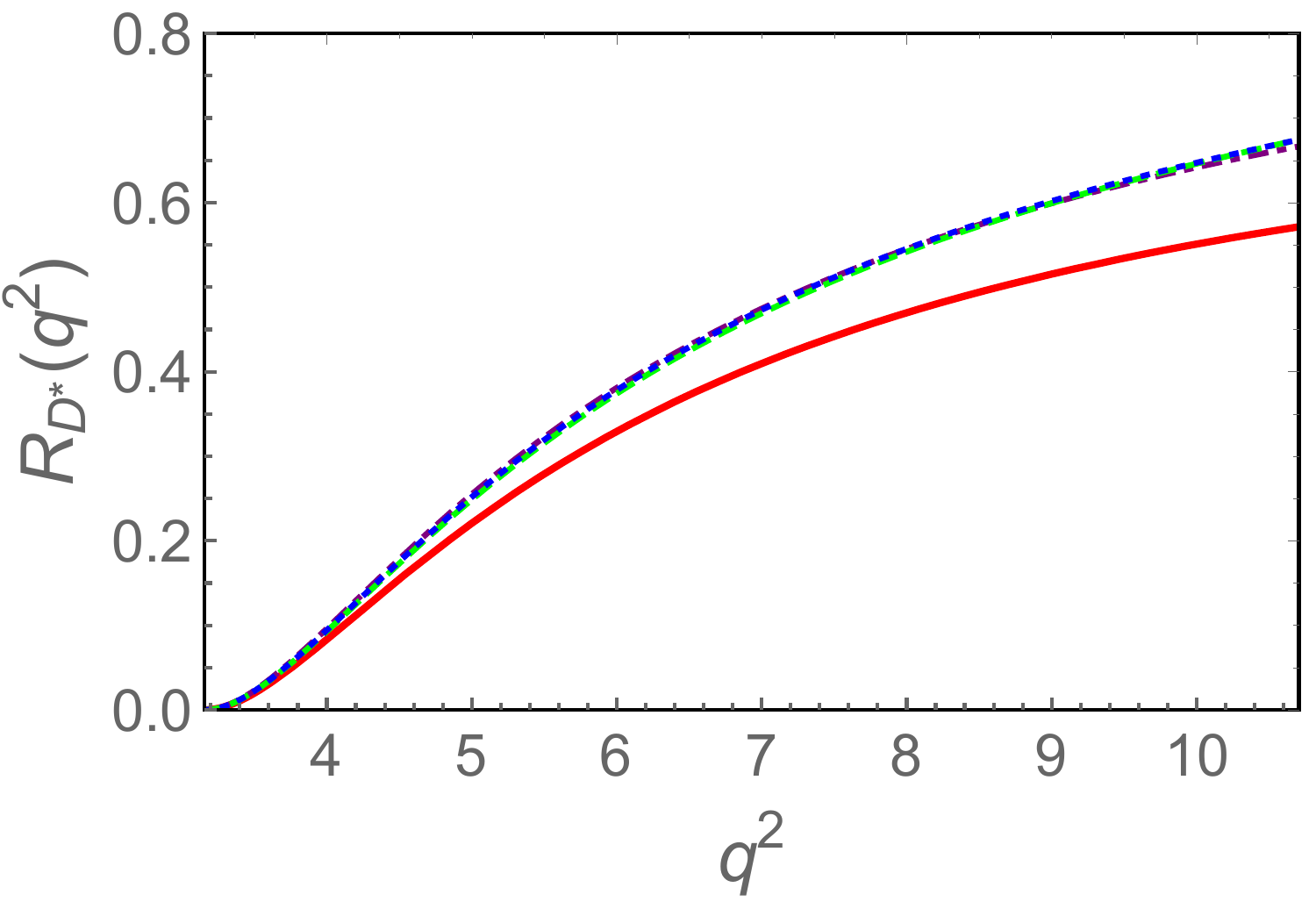}~~~
  \includegraphics[width=5cm,height=3.5cm]{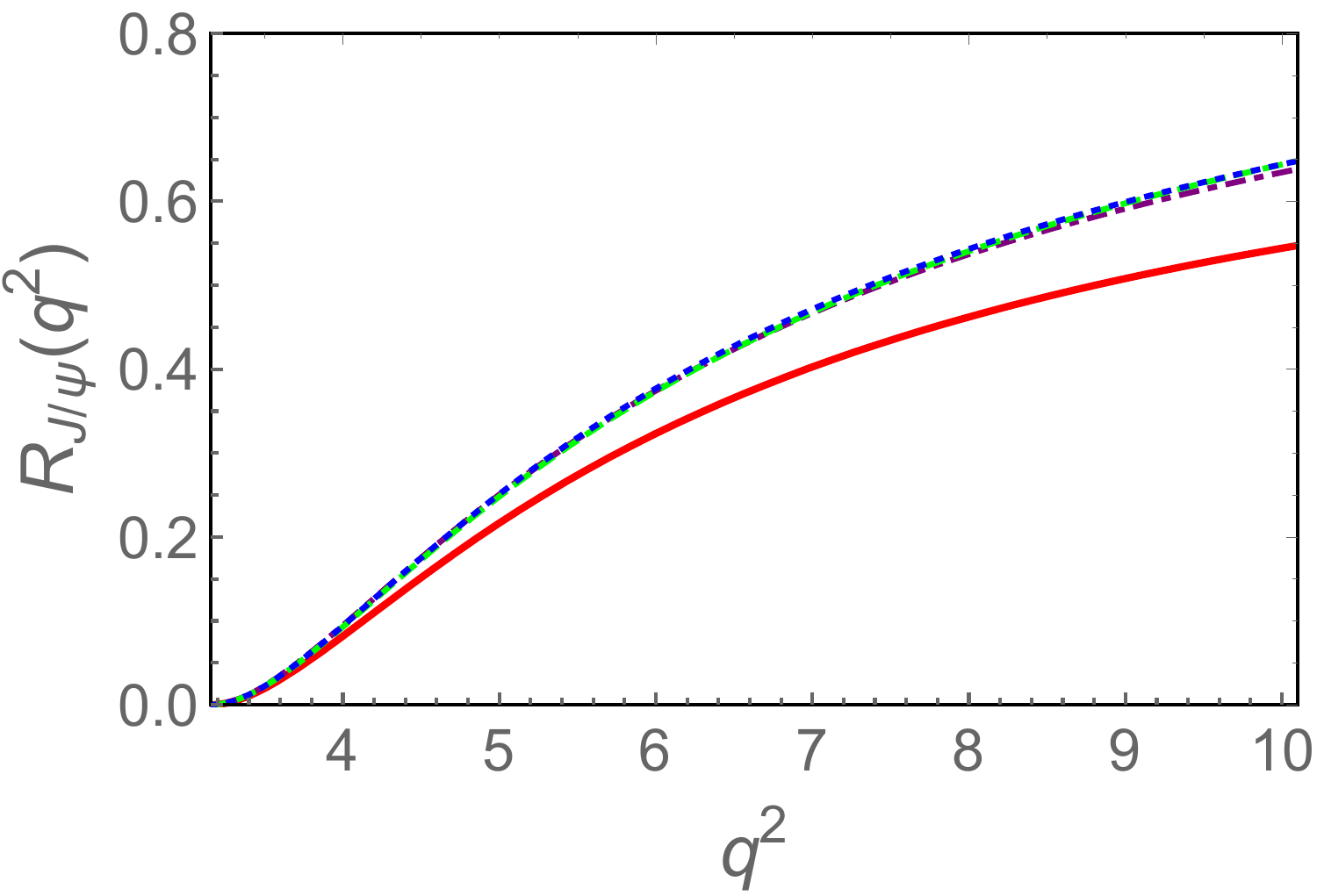}\\
  \includegraphics[width=5cm,height=3.5cm]{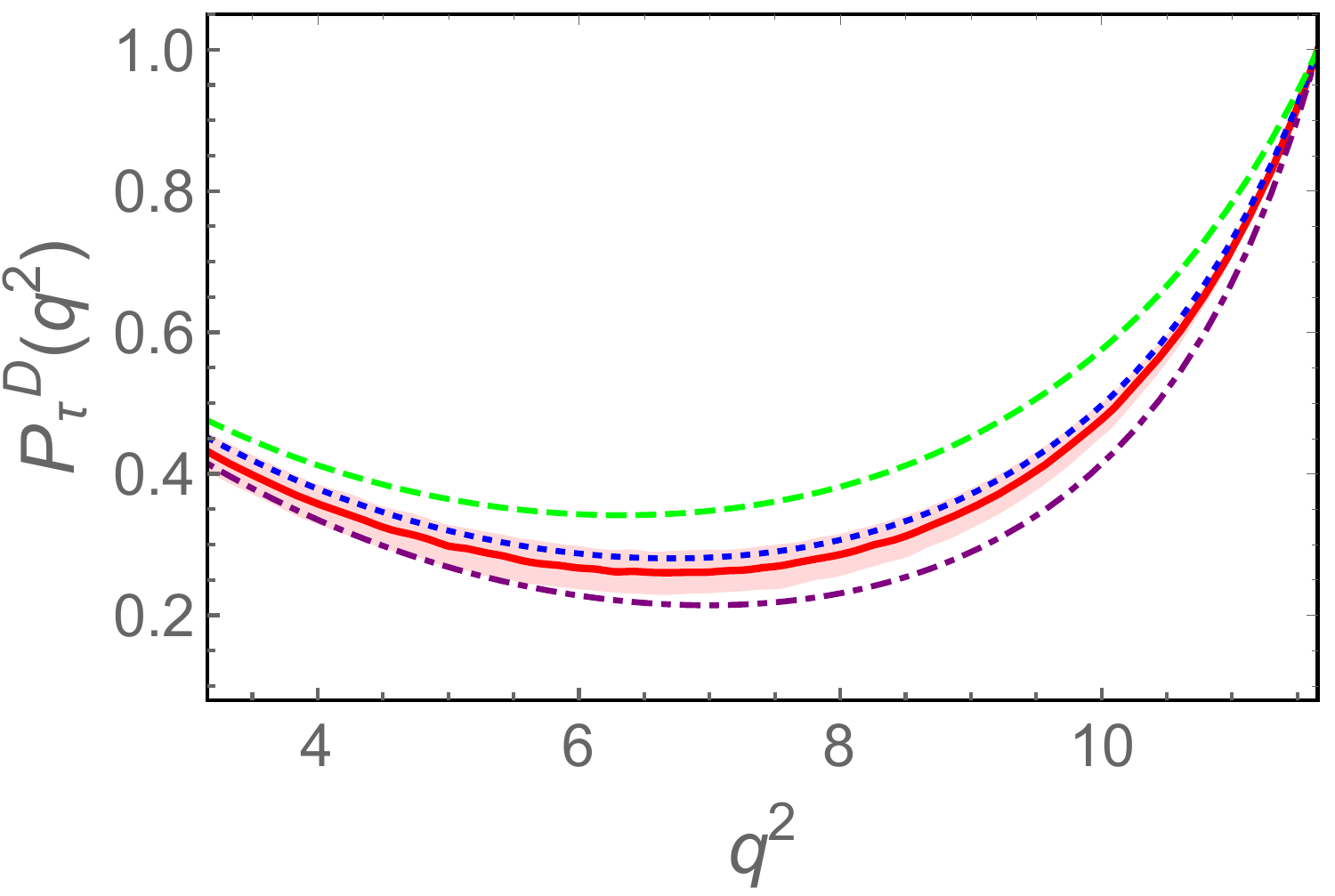}~~~\includegraphics[width=5cm,height=3.5cm]{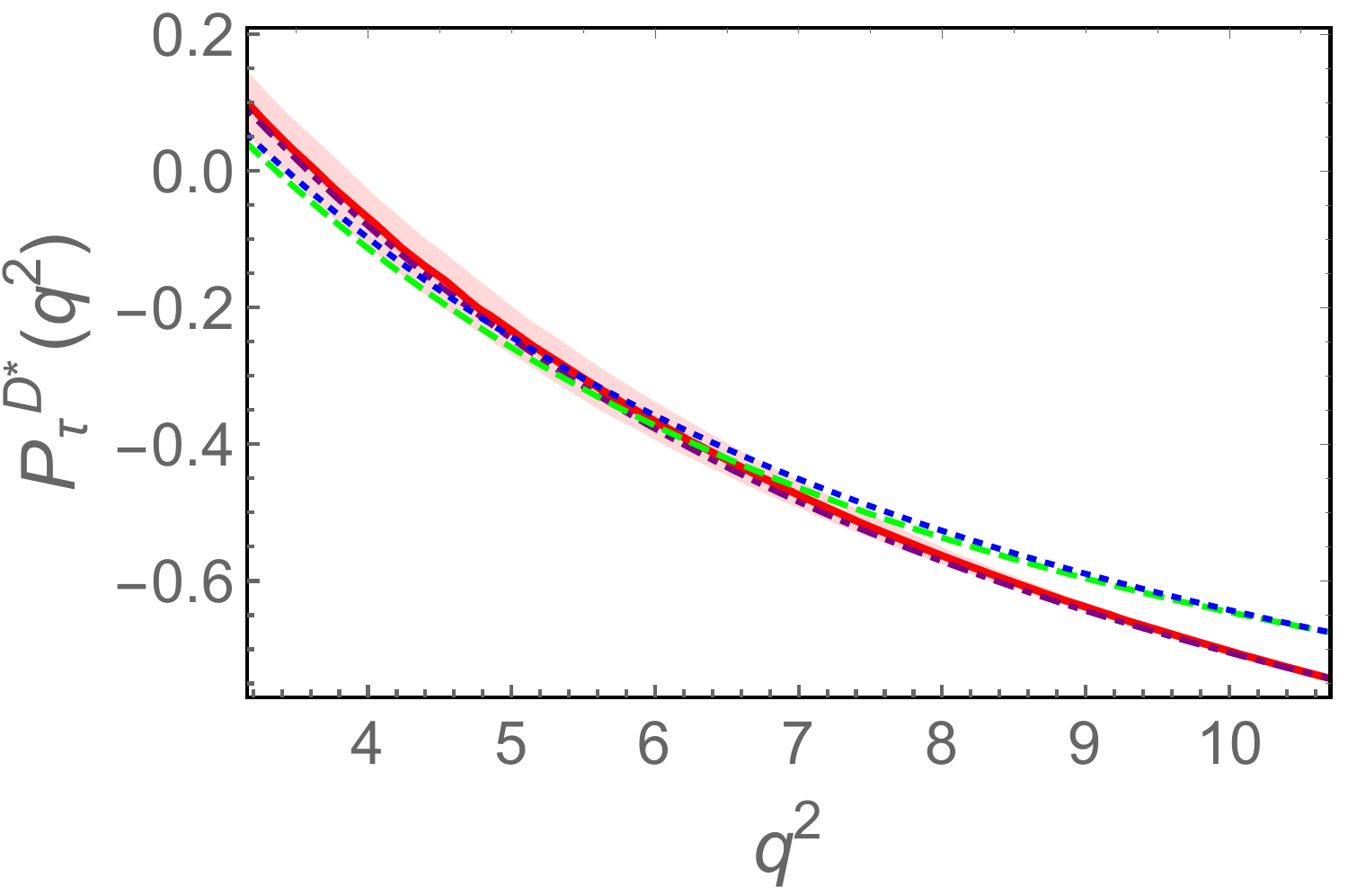}~~~
  \includegraphics[width=5cm,height=3.5cm]{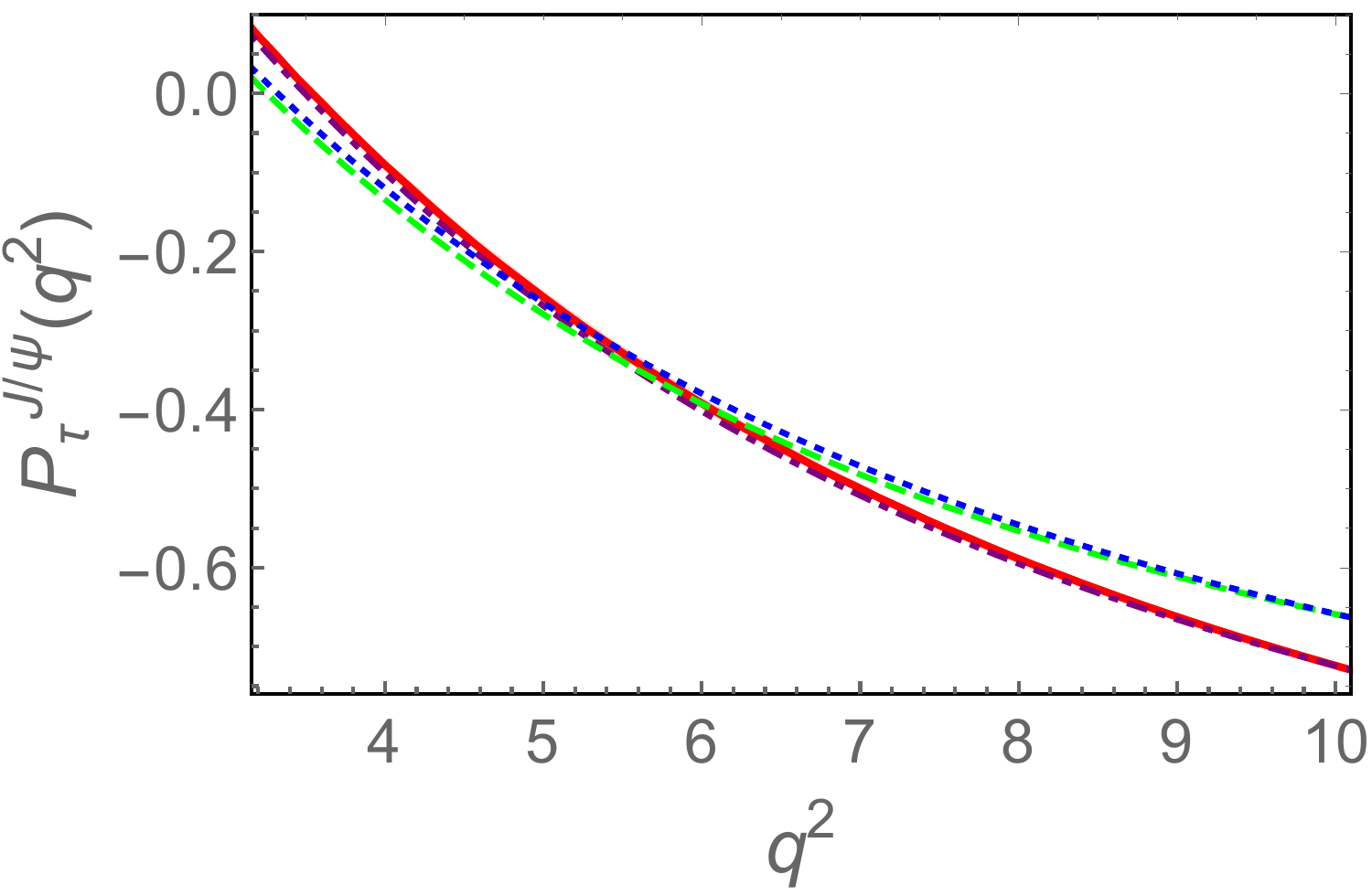}\\
  \includegraphics[width=5cm,height=3.5cm]{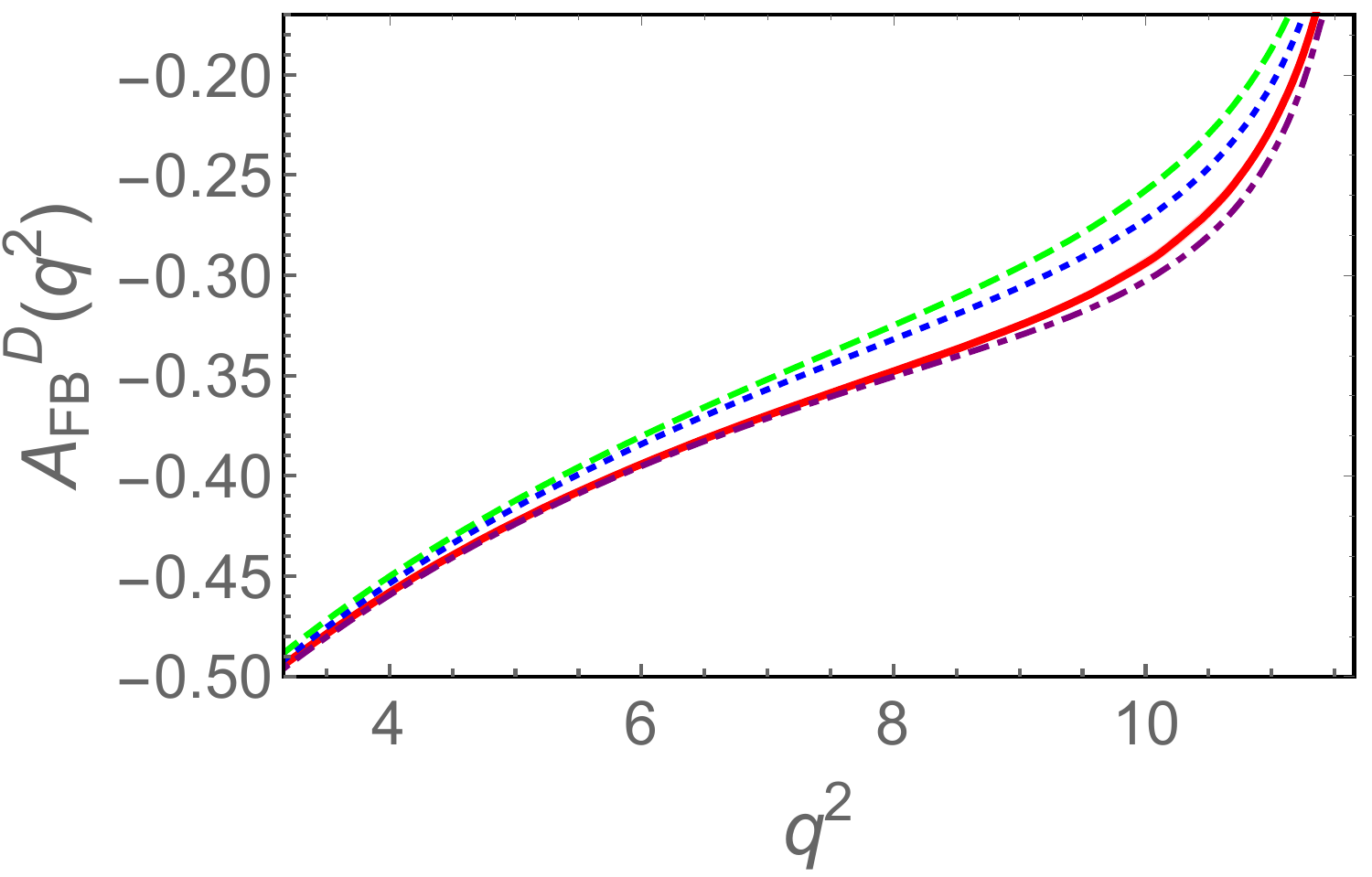}~~~\includegraphics[width=5cm,height=3.5cm]{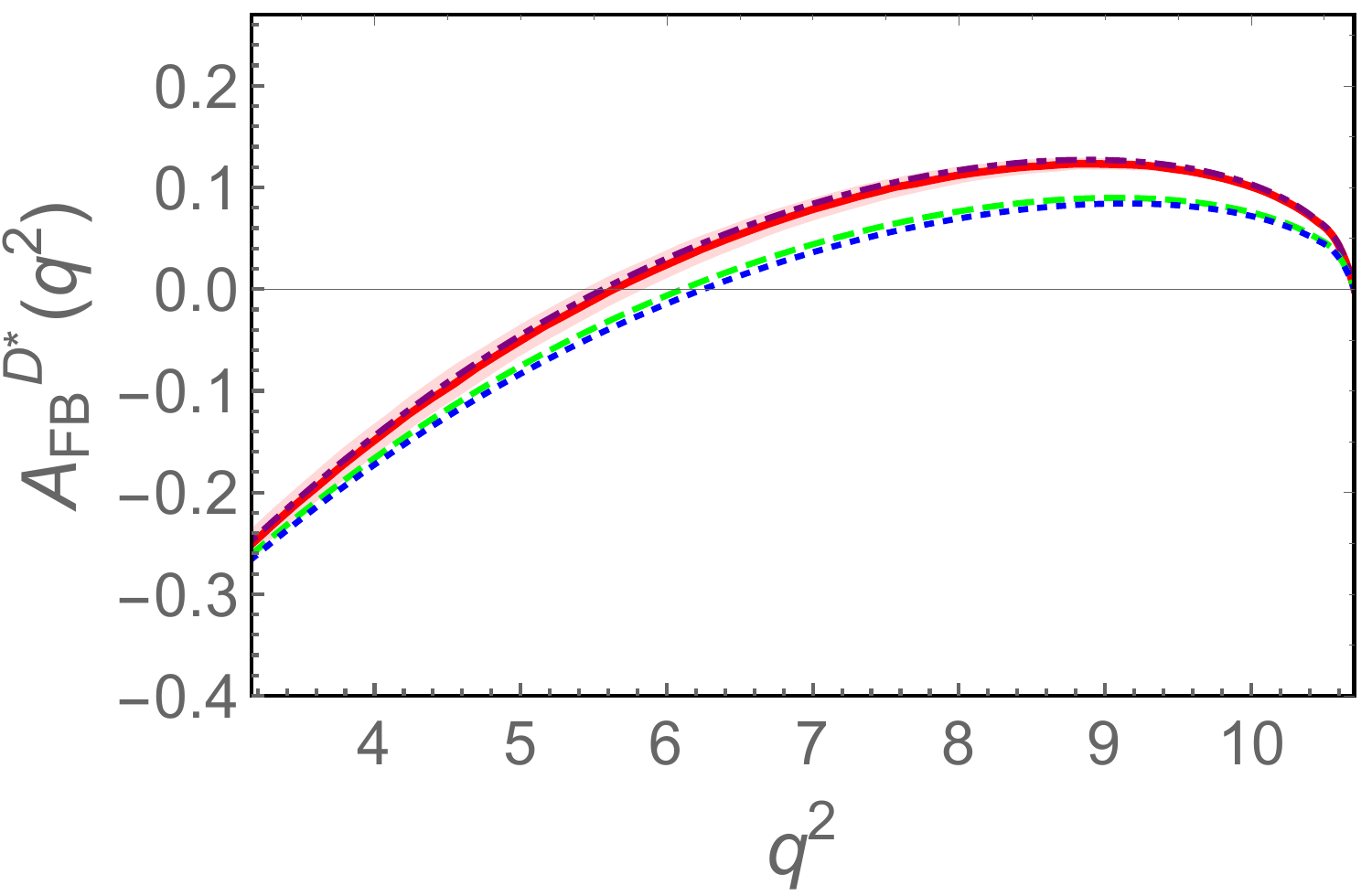}~~~
  \includegraphics[width=5cm,height=3.5cm]{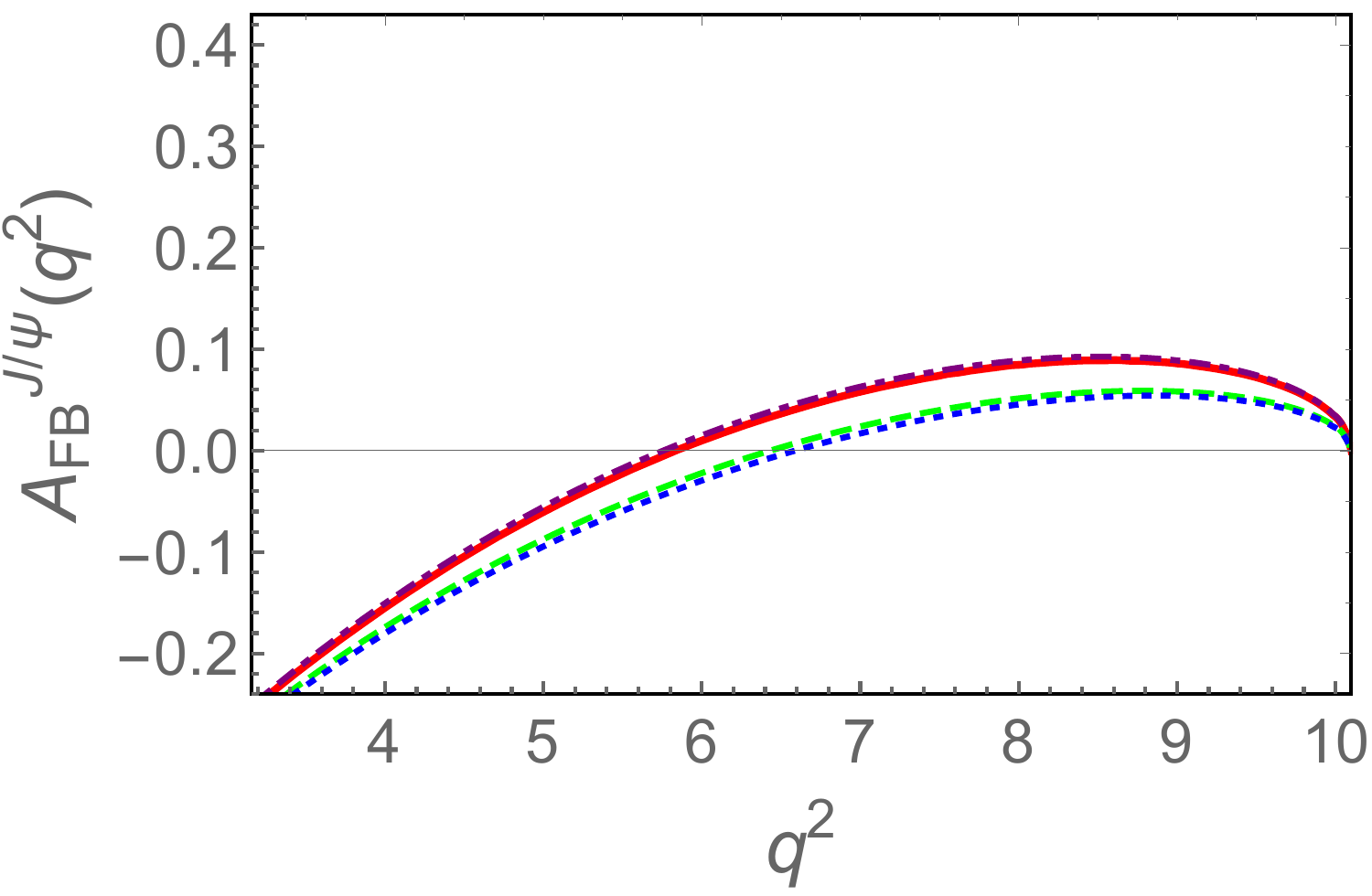}\\
  \includegraphics[width=5cm,height=3.5cm]{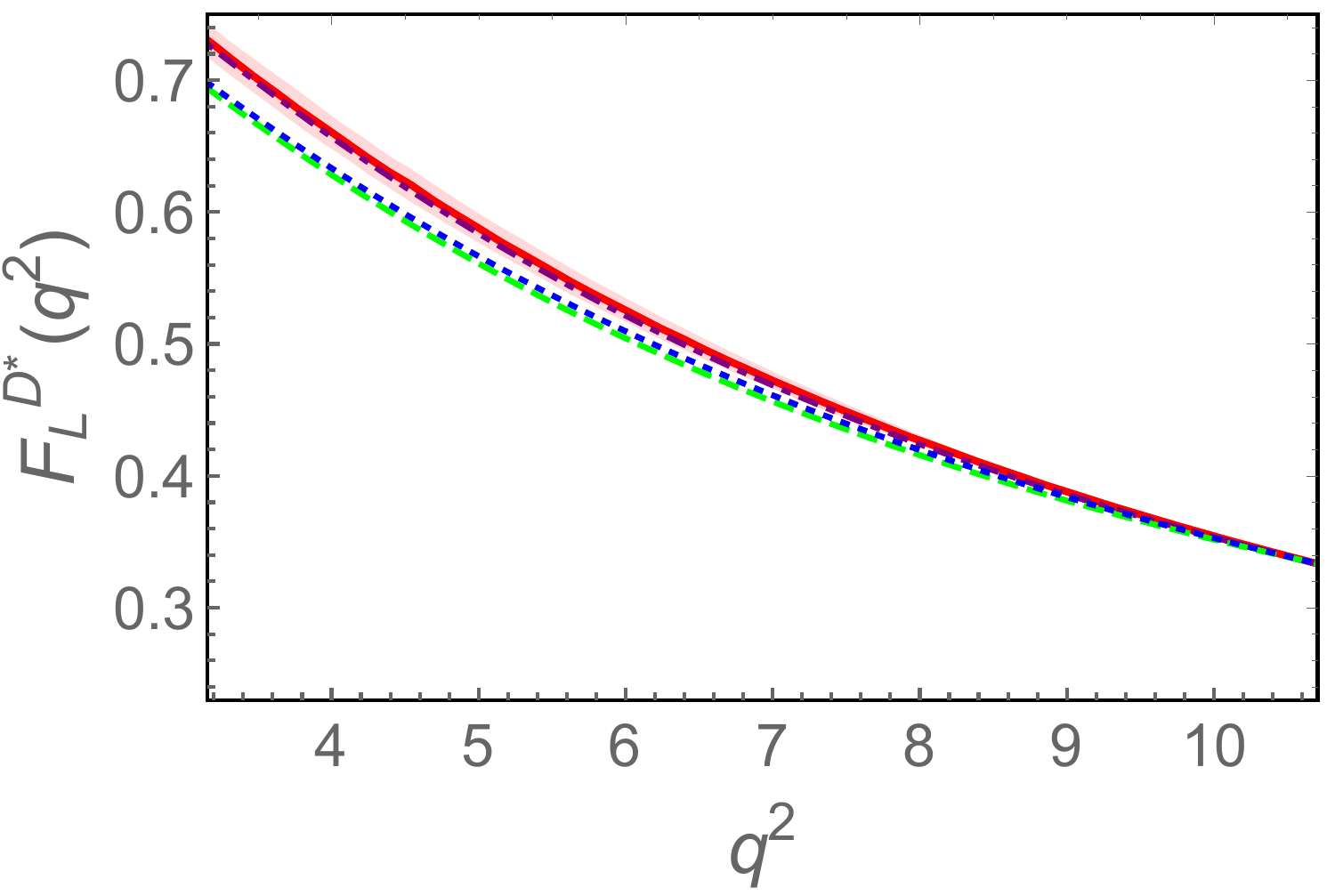}~~~\includegraphics[width=5cm,height=3.5cm]{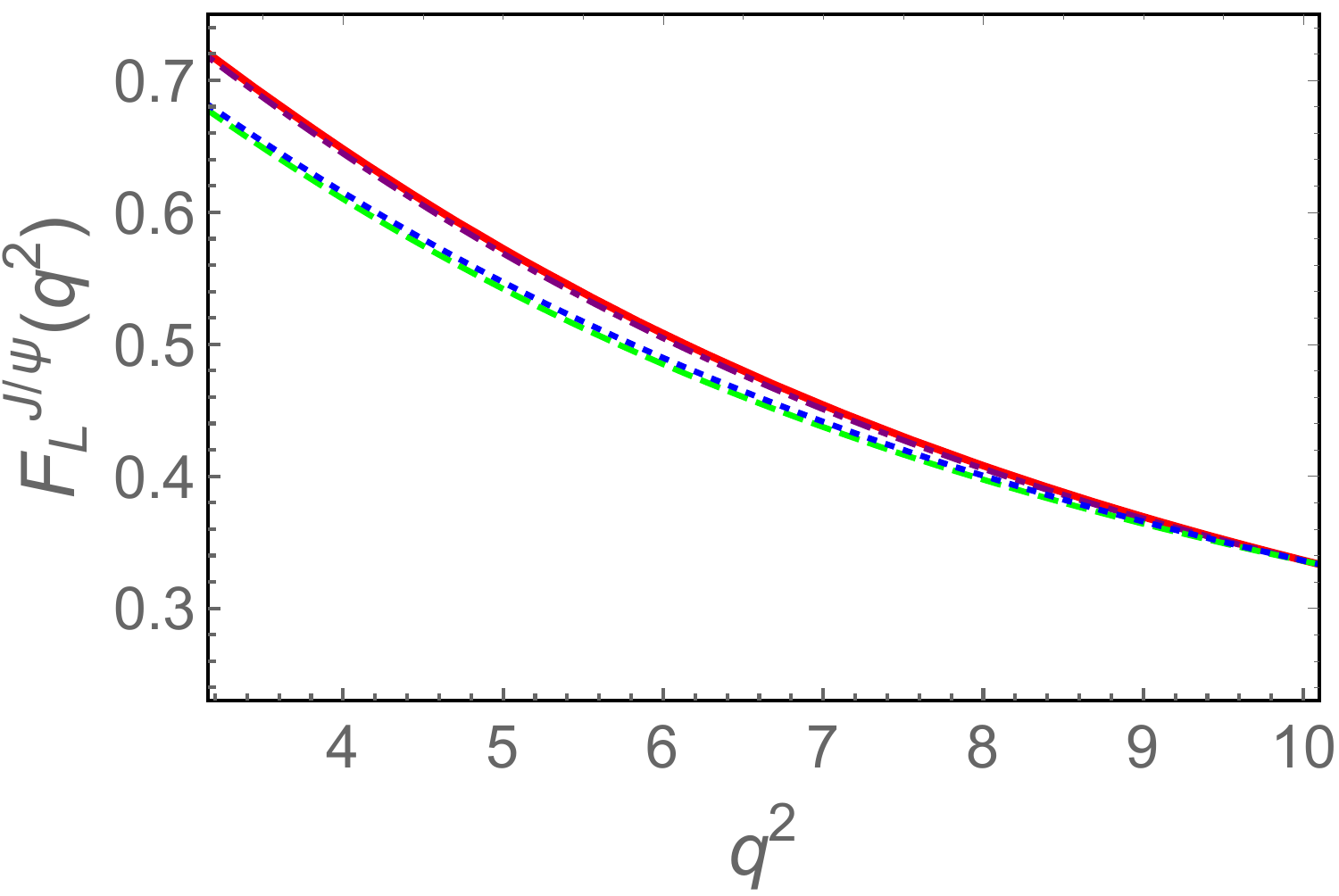}\\
  \caption{Dependence of all the observables on $q^2$ in the SM~(red-solid lines) and the NP scenarios $\epsilon_T^\tau=-0.03$~(blue dotted lines), $(\epsilon_{S_L}^\tau,~\epsilon_T^\tau)=(0.07,-0.03)$~(green dashed lines) and $(\epsilon_{L}^\tau,~\epsilon_{S_R}^\tau)=(0.08,-0.05)$~(purple dot-dashed lines). Shaded area around SM curves represent the uncertainties of the SM predictions.}\label{fig:6}
\end{figure}

\begin{table}[h!]
 \caption{Predictions in the SM and different NP scenarios for binned observables integrating over the whole kinematic regions.}\label{tab:Binned}
\begin{center}
    \begin{tabular}{cccccc}
      \hline
      \hline
     \multirow{2}{2cm}{Observables} & \multirow{2}{0.5cm}{SM} & \multirow{2}{2cm}{$\epsilon_T^\tau=-0.03$}
     & $(\epsilon_{S_L}^\tau,~\epsilon_T^\tau)$
     &$(\epsilon_{L}^\tau,~\epsilon_{S_R}^\tau)$
     & $(\epsilon_{L}^\tau,\epsilon_{T}^\tau,\epsilon_{S_L}^\tau,\epsilon_{S_R}^\tau)$\\
     & & & ~~~$=(0.07,-0.03)$~~~ & ~~~~$=(0.08,-0.05)$~~~ & ~~~$=(0.16,0.05,-0.33,0.14)$~~~\\
      \hline
      $R_D$ & $0.312_{-0.018}^{+0.019}$ & $0.303_{-0.018}^{+0.019}$ & $0.340_{-0.021}^{+0.023}$ & $0.339_{-0.018}^{+0.020}$ & $0.343_{-0.016}^{+0.017}$\\

      $P_\tau^D$ & $0.338_{-0.034}^{+0.033}$ & $0.358_{-0.034}^{+0.033}$ & $0.427_{-0.032}^{+0.032}$ & $0.288_{-0.034}^{+0.034}$ & $0.117_{-0.033}^{+0.033}$\\

      $A_{FB}^D$ & $-0.358_{-0.003}^{+0.003}$ & $-0.344_{-0.003}^{+0.004}$ & $-0.334_{-0.004}^{+0.005}$ & $-0.363_{-0.002}^{+0.002}$ & $-0.383_{-0.001}^{+0.002}$\\
      \hline
      $R_{D^*}$ & $0.253_{-0.004}^{+0.004}$ & $0.293_{-0.004}^{+0.004}$ & $0.291_{-0.003}^{+0.004}$ & $0.293_{-0.004}^{+0.004}$ & $0.297_{-0.008}^{+0.009}$\\

      $P_\tau^{D^*}$ & $-0.505_{-0.022}^{+0.024}$ & $-0.477_{-0.019}^{+0.020}$ & $-0.487_{-0.017}^{+0.019}$ & $-0.513_{-0.021}^{+0.023}$ & $-0.430_{-0.041}^{+0.042}$\\

      $A_{FB}^{D^*}$ & $0.068_{-0.013}^{+0.013}$ & $0.030_{-0.012}^{+0.012}$ & $0.038_{-0.012}^{+0.012}$ & $0.073_{-0.013}^{+0.013}$ & $0.083_{-0.016}^{+0.017}$\\

      $F_L^{D^*}$ & $0.455_{-0.008}^{+0.009}$ & $0.444_{-0.007}^{+0.008}$ & $0.440_{-0.007}^{+0.007}$ & $0.452_{-0.008}^{+0.008}$ & $0.497_{-0.014}^{+0.015}$\\
      \hline
       $R_{J/\psi}$ & $0.248_{-0.003}^{+0.003}$ & $0.291_{-0.004}^{+0.004}$ & $0.289_{-0.004}^{+0.004}$ & $0.288_{-0.004}^{+0.004}$ & $0.284_{-0.003}^{+0.003}$\\

      $P_\tau^{J/\psi}$ & $-0.512_{-0.010}^{+0.011}$ & $-0.481_{-0.008}^{+0.009}$ & $-0.490_{-0.008}^{+0.008}$ & $-0.519_{-0.010}^{+0.010}$ & $-0.453_{-0.019}^{+0.020}$\\

      $A_{FB}^{J/\psi}$ & $0.042_{-0.006}^{+0.006}$ & $0.007_{-0.006}^{+0.006}$ & $0.013_{-0.006}^{+0.006}$ & $0.046_{-0.006}^{+0.006}$ & $0.061_{-0.007}^{+0.007}$\\

       $F_L^{J/\psi}$ & $0.446_{-0.003}^{+0.003}$ & $0.434_{-0.003}^{+0.003}$ & $0.430_{-0.002}^{+0.002}$ & $0.443_{-0.003}^{+0.003}$ & $0.490_{-0.005}^{+0.005}$\\
      \hline
      \hline
    \end{tabular}
  \end{center}
\end{table}
In Fig.~\ref{fig:6}, we study the  $q^2$ spectra of $R_{D^{(*)}}$ and of a selection of polarization and angular observables~\footnote{
All of them have been defined in Sec. \ref{sec:intro}, except  the tauonic forward-backward asymmetry,
\begin{eqnarray}
A_{FB}=\frac{\int_0^1\frac{d\Gamma}{d\cos\theta}d\cos\theta-\int_{-1}^0\frac{d\Gamma}{d\cos\theta}d\cos\theta}
{\int_{-1}^1\frac{d\Gamma}{d\cos\theta}d\cos\theta},
\end{eqnarray}
which is independent of overall normalization~\cite{Alonso:2017ktd}.} showing their sensitivity to NP. We select scenarios that can be motivated by UV completions such as those involving scalar-tensor or vector-scalar combinations of operators, and we also study the tensor scenario. The values of the WCs are fixed to the results of the fits to the $R_{D^{(*)}}$ data, i.e, $\epsilon_T^\tau=-0.03$,  $(\epsilon_{S_L}^\tau,~\epsilon_T^\tau)=(0.07,-0.03)$, $(\epsilon_{L}^\tau,~\epsilon_{S_R}^\tau)=(0.08,-0.05)$. In Tab~\ref{tab:Binned} we show the results of these observables integrated over the whole kinematic region for the SM and the different NP scenarios considered. Interestingly, none of the preferred scenarios with up to two WCs can satisfactorily describe the Belle measurement of $F_L^{D^*}$ along with the experimental enhancements reported in $R_{D}$ and $R_{D^*}$.

From the plots in Fig.~\ref{fig:6} and predictions in Tab~\ref{tab:Binned}, one concludes that a clear pattern emerges in these observables for the different NP scenarios currently favored by the data, although high precision measurements will be required to discriminate among them. The most sensitive ones for this purpose turn out to be the tau polarization and forward-backward asymmetry of the $B\to D\tau\nu$ decay mode. Interestingly, with the 50 ab$^{-1}$ expected to be collected by Belle II a relative statistical uncertainty better than $\sim 10\%$ has been estimated for these observables integrated over the whole $q^2$ region~\cite{Alonso:2017ktd}.

\section{Summary and outlook}

In this work, we have studied in detail the status of the new-physics interpretations of the $b\to c\tau\nu$ anomalies after the addition of the Belle measurements of $R_{D^{(*)}}$ using the semileptonic tag and $F_L^{D^*}$ to the data set. We perform two types of fits: First, we fit with one and two parameters (Wilson coefficients) to the 2019 HFLAV average of $R_D$ and $R_D^*$ with particular attention to the evolution of the preferred scenarios with the new data and to the consistency with the upper bounds that can be derived from the lifetime of the $B_c$ meson and the $p p \to \tau_h X$+MET signature at the LHC. The main conclusion is that NP interpretations driven by left-handed currents and tensor operators are favored by the data with a significance of $\sim 3.5\sigma$ with respect to the SM hypothesis. Solutions based on pure right-handed currents remain disfavored by the LHC data while scenarios with that only have scalar contributions are in conflict with both, the LHC and the $B_c$-meson experimental inputs. In fact, the LHC upper bounds currently exclude large regions of the parameter space allowed by the $R_{D^{(*)}}$ data, and in the high-luminosity phase it should start probing all the interesting regions.

We also perform a second global fit of all the NP operators with (left-handed neutrinos) to the $R_{D^{(*)}}$ data, $R_{J/\Psi}$, $F_L^{D^*}$ and $P_\tau^{D^*}$. The main effect of the added observables, in particular of $F_L^{D^*}$, is to exclude the regions involving large values of the WCs, in complementarity with the upper LHC bounds. Otherwise, the favored regions by the global fits are equivalent to the ones resulting from the fit to  $R_{D^{(*)}}$.

A caveat to our conclusions is that the LHC bounds derived from the analysis in terms of effective operators are not applicable if the mass scale of the new mediators they correspond to is lighter than $\sim2$ TeV. Scenarios based on $S_1$ and $U_1$ leptoquarks coupled to right-handed neutrinos remain challenged by the monotau signature at the LHC except for the mass range which is being independently probed by pair-production at the LHC. A $S_1$ leptoquark producing a scalar-tensor scenario does not provide a solution as optimal as with the 2018 HFLAV average, whereas in combination with the $R_2$ leptoquark it can provide the optimal tensor scenario. The $R_2$ leptoquark alone can also explain the data successfully when the couplings take complex values and, interestingly, its detection should be at reach in the HL-LHC. Best solutions are incarnated by the $S_1$ and $U_1$ leptoquarks with pure left-handed couplings, possibly in combination with right-hand currents in the latter case.

Finally, we investigate the sensitivity of different observables to NP. We find that the tau polarization in the $B\to D\tau\nu$ decay is sensitive to the various scenarios favored by the data. Interestingly, Belle II could achieve a precision in this observable that would provide discriminating power among them.

\section{Acknowledgments}

This work is partly supported by the National Natural Science Foundation of China under Grant  No. 11735003 and by the fundamental Research Funds for the Central Universities. BG was supported in part by the US Department of Energy grant No. DE-SC0009919.  SJ was supported in part by UK STFC Consolidated Grant ST/P000819/1. JMC acknowledges support from the Spanish MINECO through the ``Ram\'on y Cajal'' program RYC-2016-20672. \\

{\bf \noindent Note added:}

While this paper was being finished different analyses of the new data set of $R_{D^{(*)}}$ have been reported~\cite{Murgui:2019czp,Bardhan:2019ljo,Asadi:2019xrc}.

\section{Appendix}

In Tables~\ref{tab:fit1corr} and~\ref{tab:fit2corr} we provide the correlation matrices for the two-parameter fits to the 2019 HFLAV average of $R_{D}$ and $R_{D^{(*)}}$, Table~\ref{tab:fit1}, and to all the observables, Table~\ref{tab:fit2}.

\begin{table}[h!]
 \caption{\label{tab:fit1corr}The $1\sigma$ uncertainty and correlation $\rho$ for two WC fits in Table~\ref{tab:fit1}.}
\begin{center}
    \begin{tabular}{ccc}
      \hline
      \hline
      ~~~~~~~~ & ~~$1\sigma$ uncertainty~~& ~~$\rho$~~\\
      \hline

      ~~$(\epsilon_{S_L}^\tau,\epsilon_T^\tau)$~~ & ~~$(\pm0.10,\pm0.01)$~~ & ~~$0.079$~~\\

      ~~$(\epsilon_{S_L}^\tau,\epsilon_{S_R}^\tau)$~~ & ~~$(\pm0.27,\pm0.25)$~~ & ~~$-0.925$~~\\

      ~~$(\epsilon_{S_R}^\tau,\epsilon_T^\tau)$~~ & ~~$(\pm0.10,\pm0.02)$~~ & ~~$0.275$~~\\

      ~~$(\epsilon_L^\tau,\epsilon_T^\tau)$~~ & ~~$(\pm0.07,\pm0.03)$~~ & ~~$0.896$~~\\

      ~~$(\epsilon_L^\tau,\epsilon_{S_L}^\tau)$~~ & ~~$(\pm0.04,\pm0.13)$~~ & ~~$-0.496$~~\\

      ~~$(\epsilon_L^\tau,\epsilon_{S_R}^\tau)$~~ & ~~$(\pm0.04,\pm0.14)$~~ & ~~$-0.733$~~\\
      \hline
      \hline
    \end{tabular}
  \end{center}
\end{table}

\begin{table}[h!]
 \caption{\label{tab:fit2corr} The $1\sigma$ uncertainty and correlation $\rho$ for two WC fits in Table~\ref{tab:fit2}.}
\begin{center}
    \begin{tabular}{ccccccc}
      \hline
      \hline
      ~~~~~~~~ & ~~$1\sigma$ uncertainty~~ & ~~$\rho$~~\\
      \hline

      ~~$(\epsilon_{S_L}^\tau,\epsilon_T^\tau)$~~ & ~~$(\pm0.10,\pm0.02)$~~ & ~~$0.070$~~\\

      ~~$(\epsilon_{S_L}^\tau,\epsilon_{S_R}^\tau)$~~ & ~~$(\pm0.26,\pm0.24)$~~ & ~~$-0.921$~~\\

      ~~$(\epsilon_{S_R}^\tau,\epsilon_T^\tau)$~~ & ~~$(\pm0.10,\pm0.02)$~~ & ~~$0.256$~~\\

      ~~$(\epsilon_L^\tau,\epsilon_T^\tau)$~~ & ~~$(\pm0.07,\pm0.03)$~~ & ~~$0.891$~~\\

      ~~$(\epsilon_L^\tau,\epsilon_{S_L}^\tau)$~~ & ~~$(\pm0.04,\pm0.12)$~~ & ~~$-0.487$~~\\

      ~~$(\epsilon_L^\tau,\epsilon_{S_R}^\tau)$~~ & ~~$(\pm0.04,\pm0.15)$~~ & ~~$-0.732$~~\\
      \hline
      \hline
    \end{tabular}
  \end{center}

\end{table}
\pagebreak

\bibliography{bctaunu}

\end{document}